\documentclass[letter,11pt]{article}

\usepackage{graphicx, epsfig, color}
\usepackage{amsmath,amssymb,hyperref}
\usepackage{caption,subcaption,graphicx}
\usepackage{titlesec}

\usepackage{xcolor}
\hypersetup{
    colorlinks,
    linkcolor={red!50!black},
    citecolor={blue!50!black},
    urlcolor={blue!50!black}
}

\def\to{\rightarrow}

\def\bi{\begin{itemize}}
\def\ei{\end{itemize}}

\def\sps1ap{SPS1a$^\prime$}
\def\c1p{C1$^\prime$}

\def\tst{\tilde t}

\def\tg{\tilde g}
\def\tnu{\tilde\nu}

\def\agt{\gtrsim}
\def\be{\begin{equation}}  
\def\ee{\end{equation}}  
\def\bea{\begin{eqnarray}}  
\def\eea{\end{eqnarray}}  
\def\beas{\begin{eqnarray*}}  
\def\eeas{\end{eqnarray*}}

\titlespacing{\section}
{0pt}{10pt}{12pt}
\titlespacing{\subsection}
{0pt}{8pt}{8pt}

\title{Lepton flavor violation from SUSY with non-universal scalars}
\begin{document}
\begin{titlepage}

\vspace{0.5cm}
\begin{center}
{\Large \bf Lepton flavor violation from SUSY with non-universal scalars
}\\ 
\vspace{1.2cm} \renewcommand{\thefootnote}{\fnsymbol{footnote}}
{\large Howard Baer$^1$\footnote[1]{Email: baer@nhn.ou.edu }, 
Vernon Barger$^2$\footnote[2]{Email: barger@pheno.wisc.edu },
and 
Hasan Serce$^2$\footnote[5]{Email: serce@ou.edu}
}\\ 
\vspace{1.2cm} \renewcommand{\thefootnote}{\arabic{footnote}}
{\it 
$^1$Department of Physics and Astronomy,
University of Oklahoma, Norman, OK 73019, USA \\
}
{\it 
$^2$Department of Physics,
University of Wisconsin, Madison, WI 53706, USA \\
}
\end{center}

\vspace{0.5cm}
%\begin{abstract}
\noindent {\bf Abstract :}
\vspace{0.4cm}

\noindent Right-handed neutrinos in supersymmetric models can act as the source of 
lepton flavor violation (LFV). 
We present experimental implications of lepton flavor-violating processes within a supersymmetric 
type-I seesaw framework in the three-extra-parameter non-universal Higgs model (NUHM3)
for large (PMNS-like) and small (CKM-like) Yukawa mixing scenarios. 
We highlight LFV predictions for the natural (low $\Delta_{\rm EW}$) portion of parameter space.
Our numerical analysis includes full 2-loop renormalization group running effects 
for the three neutrino masses and mass matrices. 
We show the projected discovery reach of various LFV experiments 
({\it i.e.} Mu2e, Mu3e, MEG-II, Belle-II), and specify regions that have 
already been excluded by the LHC searches. 
Our results depend strongly on whether one has a normal sneutrino hierarchy 
(NSH) or an inverted sneutrino hierarchy (ISH).
Natural SUSY with a NSH is already excluded by MEG-2013 results
while large portions of ISH have been or will soon be tested.
However, LFV processes from natural SUSY with small Yukawa mixing and an 
ISH seem below any projected sensitivities.
A substantial amount of the remaining parameter space of models with large PMNS-like 
sneutrino mixing will be probed by Mu2e and MEG-II experiments whereas 
small, CKM-like Yukawa mixing predicts LFV decays which can hide from LFV experiments.

\vspace*{0.8cm}

%\end{abstract}

%\keywords{flavor, neutrino, supersymmetry, mu2e, mu3e}
\end{titlepage} 

\section{Introduction}
\label{sec:intro}

Despite the lack of signals from the Large Hadron Collider 
(LHC)~\cite{Canepa:2019hph}, 
supersymmetry (SUSY) remains a very compelling model for physics Beyond the Standard Model. 
SUSY naturally solves the big hierarchy problem~\cite{witten}, 
introduces a dark matter candidate~\cite{goldberg} and explains neutrino masses 
when augmented with right-handed (RH) Majorana neutrinos~\cite{SUSYnus}.
So far, the Minimal Supersymmetric Standard Model (MSSM)~\cite{wss} has been challenged and 
yet has survived four experimental tests: 1. gauge coupling unification~\cite{gauge}, 
2. discovery of the top quark with $m_t$ within 100-200 GeV for a succesful radiative 
electroweak symmetry breaking~\cite{rewsb}, 3. the Higgs boson was discovered with $m_h\simeq 125$ GeV,
within the narrow range of MSSM allowed values~\cite{mhiggs} and 4. precision electroweak
observables-- $m_W$ vs. $m_t$ plane-- favor heavy SUSY even more than the SM~\cite{sven}. 

Regarding the neutrino sector, observations of neutrino oscillations 
confirm that neutrinos are massive particles~\cite{nureview}. 
Although individual masses of neutrinos are unknown, experiments can probe the mass gaps 
between the generations. 
Measured values of the mass gaps suggest two scenarios for the neutrino mass hierarchy: 
normal ($m_{\nu_3} > m_{\nu_2} \gtrsim m_{\nu_1}$) and inverted ($m_{\nu_3} < m_{\nu_1} \lesssim m_{\nu_2}$). 
A global analysis based on oscillation and non-oscillation data within a three-neutrino 
framework performed by Capozzi {\it et al.}~\cite{Capozzi:2017ipn} with $3\sigma$ deviations 
gives~\cite{Tanabashi:2018oca}
\be
\Delta m_{21}^2 =(7.37^{+0.59}_{-0.44})\times \: 10^{-5}{\rm eV}^2,  \qquad
\sin \theta_{12}=0.297^{+0.57}_{-0.47}
\ee
and
\vspace*{-0.2cm}
\begin{align*}
&{\rm normal \:\: hierarchy \:(NH)}: &   &{\rm inverted \:\: hierarchy \:(IH)}: \\
\Delta m_{31}^2 & =(2.56^{+0.13}_{-0.11})\times \: 10^{-3}{\rm eV}^2 & 
\Delta m_{23}^2 & =(2.54^{+0.11}_{-0.11})\times \: 10^{-3}{\rm eV}^2 \\
\sin \theta_{13} & =0.0215 \: \pm 0.25 & 
\sin \theta_{13} & =0.0216 \: \pm 0.26 \\
\sin \theta_{23} & =0.425^{+0.19}_{-0.044} & 
\sin \theta_{23} & =0.589^{+0.047}_{-0.205}
\label{osc}
\end{align*}
where $\Delta m_{ij}^2\equiv m_i^2-m_j^2$. 
Analyses based on recent observations from IceCube and NOvA favor the normal neutrino mass 
hierarchy~\cite{Aartsen:2019eht,NOvA:2018gge}.

In a more recent study~\cite{Deppisch:2018flu}, it has been shown that current neutrino data 
can only have a good $\chi^2$ fit for $SO(10)$-based models when SUSY threshold corrections 
are included. 
The analysis is carried out using PMNS mixing and the best-fit points give:
\begin{table}[h!]
\begin{tabular}{llll}
$m_{\nu_1}=0.0018-0.0024$ eV, & $m_{\nu_2}=0.0088-0.009$ eV, & $m_{\nu_3}=0.0503$ eV & (NH)
\end{tabular}
\end{table}\\
for neutrino masses with a normal hierarchy and where the range arises from 
different $\tan \beta$ values. 
Using the $\Delta m_{ij}^2$ values given above, a lower limit for the sum of the masses of 
three neutrino generations ($\sum_{i=1,3}m_{\nu_i}$) can be estimated by setting the lightest 
neutrino mass to $zero$. 
Cosmological observations can provide an upper bound on the sum. 
Depending on the model and the dataset considered for the analysis, 
$\sum^{\max}_{\nu_i}$ can range from 0.18 eV to 1.08 eV at $2 \sigma$ level~\cite{Capozzi:2017ipn}. 

The current sensitivity of precision LFV measurements already probes certain regions of 
SUSY parameter space, especially the constrained MSSM (cMSSM or mSUGRA)~\cite{cmssm} with light scalar masses and right-handed neutrinos (RHN). 
The MEG Collaboration reported an upper bound for the branching fraction of the process 
$\mu \to e \: \gamma$: BF$(\mu \to e \: \gamma) < 4.2 \times 10^{-13}$
 at 90\% CL~\cite{TheMEG:2016wtm}.
Given the limits, sub-TeV cMSSM+RHN models that assume small CKM-like and large 
PMNS-like Yukawa mixings are excluded-- in agreement with the LHC searches. 
The projected sensitivity of MEG-II should lower the BF limit by another order of 
magnitude~\cite{Renga:2018fpd}. 
For $\mu\rightarrow e$ conversion rate (CR) in nuclei, 
the Mu2e experiment has a projected sensitivity in the range of $10^{-16}-10^{-17}$. 
With such a projected sensitivity, Mu2e will be probing the parameter space of MSSM models 
with non-universal Higgs soft masses~\cite{Calibbi:2012gr,Bora:2014mna}.

Even though precision experiments cannot be thought of as a replacement for LHC, 
they can be complementary. 
Positive results for sparticle searches at the LHC along with LFV search results 
would narrow down many BSM theories that are presently compatible with observations. 
Null results from precision experiments can only constrain models with 
specific assumptions and so are insufficient to give universal bounds on sparticle masses. 
Similarly, MSSM+RHN models include PMNS neutrino mixing to explain neutrino oscillations
(which is distinct from the CKM-like or PMNS-like Yukawa mixing to be introduced shortly). 
In addition, the predicted LFV observables might vary by a few orders of 
magnitude depending on the choice of the RH neutrino mass spectrum 
for neutrinos (see Fig.~\ref{fig:d2}).

In this paper, we study the predictions of SUSY models with non-universal Higgs parameters 
augmented with three RH neutrino superfields $\hat{N}^c_i$. 
Motivated by the normal and inverted mass hierarchies of neutrino masses, 
we study the NUHM3 model~\cite{nuhm2} with scalar mass relations $m_0(1) \simeq m_0(2) \neq m_0(3)$. 
We investigate the parameter space where LHC searches, 
dark matter detection experiments and LFV observables can be complementary. 
For each of the different neutrino Yukawa coupling scenarios and mass hierarchies,
we take the third generation neutrino mass to be $\sim 0.05$ eV and only accept solutions 
with $m_{\nu_1}$ and $m_{\nu_2}$ within the limits given in Ref.~\cite{Capozzi:2017ipn} 
for both inverse and normal neutrino mass hierarchies.

We also show predictions for the well-motivated NUHM3 model for points that exhibit 
electroweak naturalness: $\rm \Delta_{\rm EW}\leq 30$. 
The naturalness measure $\rm \Delta_{\rm EW}$~\cite{Baer:2012up} only requires that 
weak scale contributions to the $Z$-boson mass, $m_Z$, should be comparable to or smaller than 
$m_Z$. 
From the minimization conditions for the MSSM Higgs potential\footnote{We use $\bar{\mu}$ for the SUSY $mu$ term to distinguish it from $\mu$ for muon.}:
\vspace*{0.1cm}
\be \frac{m_Z^2}{2} = \frac{m_{H_d}^2 +
\Sigma_d^d -(m_{H_u}^2+\Sigma_u^u)\tan^2\beta}{\tan^2\beta -1} -\bar{\mu}^2
\label{eq:mzs}
\ee 
electroweak naturalness is defined as ${\rm \Delta_{\rm EW}} = max_i |C_i| / (m_Z^2/2)$ 
where the $C_i$s represent the various terms on the right-hand side of Eq.~(\ref{eq:mzs}). 
Here, $\tan\beta =v_u/v_d$ is the ratio of Higgs field vevs with $m_{H_u}^2$ and $m_{H_d}^2$
the Higgs soft breaking masses and the $\Sigma_u^u$ and $\Sigma_d^d$ terms contain
over 40 radiative corrections (for expressions, see Ref.~\cite{Baer:2012cf}).

One can see that $\rm \Delta_{\rm EW}<30$ is a conservative condition which 
accomodates 3.3\% or less finetuning and allows $\bar{\mu}$ up to $\sim$ 360 GeV. 
The condition for naturalness puts strong bounds on stop and gluino masses~\cite{Baer:2018hpb}. 
For various natural MSSM models, and independent of the neutrino sector, 
upper limits have been calculated as  $m_{\tst_1} \lesssim 3.5$ TeV and $m_{\tg} \lesssim 6$ TeV 
(with the exception of the natural anomaly mediated SUSY breaking model (nAMSB)~\cite{nAMSB} 
where gluino mass can reach up to 9 TeV). 
Such models will only be partially probed by high luminosity (HL)-LHC~\cite{CidVidal:2018eel}. 
For natural SUSY, the NUHM3 allows for heavier $1^{st}$ and $2^{nd}$ generation sparticles 
ranging up to $30-40$ TeV without violating the naturalness condition that 
${\rm \Delta_{\rm EW}}<30$~\cite{Baer:2017uvn}.

In $R$-parity conserving SUSY models, the lightest SUSY particle (LSP) is expected to be stable. 
The neutralino LSP has been confronted by direct and indirect dark matter (DM) searches~\cite{Baer:2016ucr}. 
In this paper, we consider a neutralino LSP with a thermal abundance less than the 
measured value, $\Omega^{\rm th} h^2 \leq 0.12 $.
In this case, the neutralino makes up only a fraction of the total dark matter density whilst 
the remainder might be composed of axions. 
In such a scenario, the neutralino abundance can also be augmented by late-decaying relics, 
such as axinos and saxions~\cite{boltz}. 
Although direct DM searches probe deeper in SUSY parameter space, indirect searches 
from FERMI-LAT and other experments limit the composition of mixed dark matter 
more strongly~\cite{Serce:2017vtk} since the WIMP indirect detection rates are 
rescaled by $\xi^2=(\Omega h^2/0.12)^2$.
 
In Sec.~\ref{sec:seesaw}, we review the SUSY seesaw mechanism while in Sec.~\ref{sec:flav}
we discuss possible LFV processes. 
In Sec~\ref{sec:gl}, we describe the model parameter space and present the  results for NUHM3+RHN. 
We summarize projected discovery limits of future LFV experiments in Sec.~\ref{sec:conclude}.

\section{Seesaw Mechanism in MSSM}
\label{sec:seesaw}

The seesaw mechanism is one of the most compelling ways to generate the observed 
neutrino masses~\cite{seesaw}. 
Implications of supersymmetric models with various seesaw mechanisms have been extensively 
studied in the literature~\cite{Hisano,Masiero,Moroi:2013vya,Hirsch:2012ti,Barger:2009gc}. 
In supersymmetric models augmented with right-handed neutrinos, loop effects can 
enhance off-diagonal elements of the slepton mass matrix and hence generate sizeable 
lepton flavor violation effects which results in low energy LFV observables~\cite{Hisano,Borzumati:1986qx}.

In the supersymmetric type-I seesaw, the MSSM superpotential is of the form
\be
\hat{f}_{\rm MSSM}=\bar{\mu}{\hat H}_u^a{\hat H}_{da}+ \sum_{i,j=1,3}\left[
({\bf f}_u)_{ij}\epsilon_{ab}\hat{Q}^a_i\hat{H}_u^b\hat{U}^c_j +
({\bf f}_d)_{ij}\hat{Q}^a_i\hat{H}_{da}\hat{D}^c_j +
({\bf f}_e)_{ij}\hat{L}^a_i\hat{H}_{da}\hat{E}^c_j \right] .
\label{eq:SPot}
\ee
It is augmented by an additional set of terms containing a right-handed neutrino $\hat{N}^c$~\cite{wss}:
\be
\hat{f}=\hat{f}_{\rm MSSM}+\sum_{i,j=1,3}\left[\frac{1}{2}({\bf M}_N)_{ij}\hat{N}^c_i\hat{N}^c_j +
({\bf f}_\nu)_{ij}\epsilon_{ab}\hat{L}^a_i\hat{H}_u^b\hat{N}^c_j \right]
\ee
where ${\bf M}_N$ is the Majorana mass matrix for the heavy right-handed neutrinos. 

Since the entirety of neutrino mixing data is insufficient to fix all the elements of ${\bf f}_\mu$, it is common to assume in addition some GUT-motivated
ans\"atze which can span the realm of possibilities~\cite{Calibbi:2012gr}.
Three common scenarios include the following:
\bi
\item scenario $\#1$: $({\bf f}_\nu)_{ij}=({\bf f}_u)_{ij}$\ \ \ 
small mixing (CKM-like),
\item scenario $\#2$: $({\bf f}_\nu)_{ij}=({\bf f}_u)^{\rm diag}_{ii} \: {\rm \bf U}^{\rm PMNS}_{ij}$\ \ \ large mixing (PMNS-like) and
\item scenario $\#3$: $({\bf f}_\nu)_{ij}=3 \times ({\bf f}_u)_{ij}$
\ei
where the conditions are imposed at the GUT scale $m_{\rm GUT}$~\cite{Calibbi:2012gr}. 
One simple GUT-motivated scenario is where Higgs superfields reside in 
10-plets; in that case, then scenario \#1 is the trivial choice whereas 
for the fields that reside in 126-plets, 
then scenario \#3 should be considered for models with CKM-like neutrino couplings. 
For a given light neutrino spectrum, in scenario \#3 Majorana neutrinos should be chosen 
a factor of $\sim 3^2=9$ heavier compared to the scenario $\#1$. 
Hence, LFV observables are expected to be diminished by a factor of $\propto \log 9$ and 
enhanced by a factor of 9 due to the squared neutrino coupling in calculation: 
$(3 \times f_{\nu})^2$. 
For simplicity, we only show results for scenario \#1 (small mixing, CKM-like) 
and scenario \#2 (large mixing, PMNS-like). 
While the above GUT scenarios are likely to live in the 
swampland~\cite{Vafa:2005ui} of string inconsistent models 
(due to the presence of large Higgs multiplets~\cite{Halverson:2018xge}), 
the large and small mixing cases can also be present in string-derived models
(such as heterotic models compactified on certain orbifolds~\cite{Buchmuller:2007zd}) 
which only include the MSSM+RHN fields in the low energy spectrum, 
and where GUT multiplets arise from {\it local} grand 
unification~\cite{Buchmuller:2005sh} present at orbifold fixed points.

Above the highest see-saw scale set by the heaviest Majorana neutrino, 
denoted by $N_3$\footnote{We consider a model with three non-degenerate right-handed neutrinos.}, 
the light neutrino mass matrix can be written as:
\be
({\bf m}_\nu)_{il}= - (v\cdot \sin\beta)^2 ({\bf f}_{\nu}^T)_{ij} ({\bf M}_N^{-1})_{jk} 
({\bf f}_\nu)_{kl}
\label{eq:lneut}
\ee
where $v\equiv\sqrt{v_u^2+v_d^2}\simeq 174$ GeV is the combined vacuum expectation value of the 
Higgs bosons in the MSSM and where $\beta=\arctan(v_u/v_d)$.
The magnitudes of the neutrino Yukawa couplings $({\bf f}_\nu)_{ij}$ and 
Majorana mass matrix $({\bf M}_N)_{jk}$ in Eq.~(\ref{eq:lneut}) are functions of the energy scale $Q$. 
When the heaviest Majorana neutrino is integrated out, an effective dimension-5 
neutrino mass operator is generated :
\be
\hat{f}\ni  \frac{1}{2}({\bf \kappa})_{il}\epsilon_{ab} \hat{L}^a_i\hat{H}_u^b
\epsilon_{df} \hat{L}^d_l\hat{H}_u^f
\ee
where the coefficient ${\bf \kappa}$ of the operator  can be determined by the matching condition at the decoupling scales~\cite{Antusch:2005gp}:
%\be
%(\kappa)_{il} \vert _{M^-_{N_k}} = (\kappa)_{il} \vert _{M^+_{N_k}} + ({\bf f}_\nu)_{ij}^T (M^{-1}_{N_k})_{jk} ({\bf f}_\nu)_{kl} \vert _{M^+_{N_k}}
%\ee
\be
(\kappa)_{il} \vert _{M^-_{N_k}} = (\kappa)_{il} \vert _{M^+_{N_k}} + 
({\bf f}_{\nu}^T)_{ik} (M^{-1}_{N_k}) ({\bf f}_\nu)_{kl} \vert _{M^+_{N_k}}
\ee
where $k=1 ,2 ,3$ and where $M_{N_k}^+$ ($M_{N_k}^-$) denotes the value as 
the scale of decoupling of the $k$-th generation RHN $M_{N_k}$ is approached
from above (below). 
At $Q=M_{N_k}$, $\hat{N}_k$ is integrated out hence heavy neutrinos decouple from the theory. 
For $Q>M_{N_3}$, $\kappa=0$ and RG evolution only governs ${\bf f}_\nu$ and $M_N$. 
After the heaviest neutrino is decoupled, the neutrino mass matrix can becomes:
\be
({\bf m}_\nu)_{il}= - \frac{1}{2} (v\cdot \sin\beta)^2 (\kappa_{il}+
({\bf f}_{\nu}^T)_{ij} ({\bf M}_N^{-1})_{jk} ({\bf f}_\nu)_{kl}).
\label{eq:lneutdec}
\ee
After all the RH neutrinos are decoupled, the RGEs only govern $\kappa$ 
from the scale of the lightest heavy neutrino mass, $M_{N_1}$, all the way down to the 
electroweak breaking scale. 
For $Q<M_{N_1}$, the neutrino mass matrix is given by:
\vspace*{-0.5cm}

\be
({\bf m}_\nu)_{il}= - \frac{1}{2} (v\cdot \sin\beta)^2 \kappa_{il} ,
\label{eq:lneutw}
\ee
and physical light neutrino masses can be obtained by diagonalizing $({\bf m}_\nu)_{il}$.
\begin{figure}[h!]
\begin{center}
\includegraphics[width=.5\textwidth]{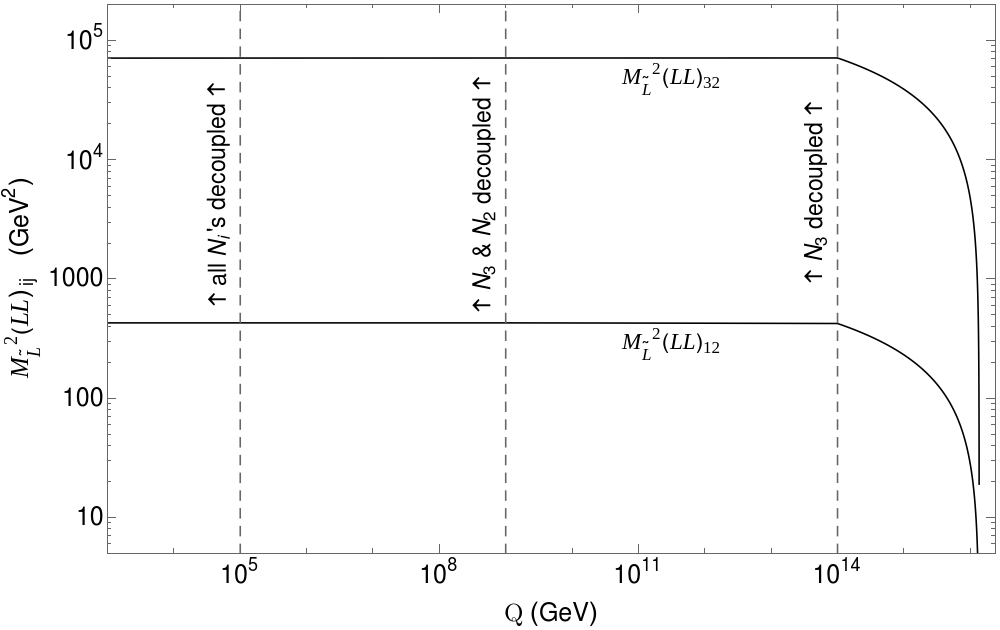}
\caption{Evolution of the off-diagonal slepton mass matrix terms 
$m^2_{\widetilde L}(LL)_{12}$ and $m^2_{\widetilde L}(LL)_{32}$ with the energy scale $Q$ from 
$m_{\rm GUT}$ to $m_Z$ with $M_{N_{1,2,3}}=10^{5,9,14}$ GeV (denoted by vertical dashed lines).
\label{fig:evolml2}}
\end{center}
\end{figure}
\vspace*{-0.5cm}

In Fig.~\ref{fig:evolml2}, we show the evolution of two off-diagonal slepton mass matrix terms 
with the energy scale $Q$ from $m_{\rm GUT}$ to the weak scale using matching conditions 
at the thresholds and the small mixing (CKM-like) neutrino Yukawa couplings at $m_{\rm GUT}$. 
The RH neutrino masses are shown by the dashed lines to be $M_{N_{1,2,3}}=10^{5,9,14}$ GeV. During the running, the off-diagonal terms receive contributions 
proportional to $Y_t^2 V^{\rm CKM}_{31}V^{\rm CKM}_{32}$ for $Q > M_{N_3}$, to $Y_c^2 V^{\rm CKM}_{21}V^{\rm CKM}_{22}$ for $M_{N_3} > Q > M_{N_2}$ 
and to $Y_u^2 V^{\rm CKM}_{11}V^{\rm CKM}_{12}$ for $M_{N_2} > Q > M_{N_1}$ 
for the CKM-like mixing as shown in the figure.
Since the $3^{rd}$ generation up-type Yukawa coupling dominates over $1^{st}$ and $2^{nd}$ 
generations, the main contribution from the RGEs is during $Q > M_{N_3}$. 
For $Q<M_{N_3}$, the RGE running effects on LFV observables are highly suppressed 
due to the smallness of up and charm quark Yukawa couplings.
\begin{figure}[t]
  \centering
  {\includegraphics[width=.48\textwidth]{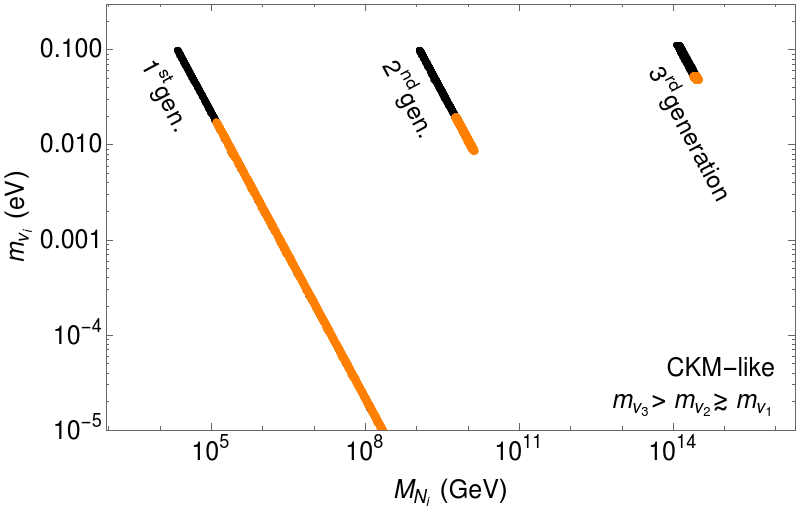}}\quad
  {\includegraphics[width=.48\textwidth]{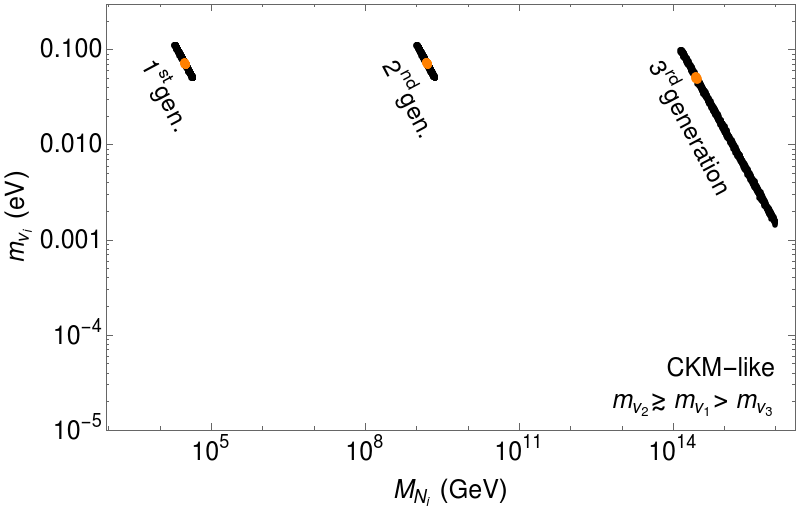}}\\ 
\caption{Dependence of light neutrino masses on heavy neutrino masses for normal hierarchy (left) and 
inverse hierarchy (right). The parameter space compatible with $m_{\nu_3}\simeq 0.05$ eV are colored in orange.}  
\label{fig:nmass}
\end{figure}
\vspace*{-0.75cm}

Using the method described above, the computed light neutrino masses ($m_{\nu_i}$)  
as a function of heavy neutrino masses ($M_{N_i}$) are shown in Fig.~\ref{fig:nmass} 
for both normal (NH) and inverted (IH) light neutrino mass hierarchies 
using the small CKM-like neutrino mixings (scenario $\#$1). 
The parameter space of neutrinos studied in this paper is shown in orange.
For values $m_{\nu_3} < 10^{-3}$ eV, $M_{N_3}$ must be close to $m_{\rm GUT}$ so we show $M_{N_3}$ 
up to $10^{16}$ GeV for the inverted hierarchy case. 
The summed neutrino masses are bounded above by  $\max (\sum {\nu_i})=0.31\ (0.32)$ eV 
for normal (inverted) hierarchy from the analysis of 
the $\Lambda {\rm CDM} + \sum + A_{\rm lens}$ model using the \textquoteleft Planck ${}_{\rm TT, TE, EE}$+ $\tau_{\rm HFI}$ + BAO\textquoteright\ dataset~\cite{Capozzi:2017ipn}.
%In the case of scenario $\#2$  where $({\bf f}_\nu)_{ij} = 3 \times ({\bf f}_u)_{ij}$, neutrino masses are multiplied by $3^2=9$ for a given set of RH neutrino masses $M_{N_i}$.

The size of the off-diagonal elements in the slepton mass matrix are mostly sensitive to 
$\log (m_{\rm GUT}/M_{N_3})$. 
As further discussed in the next section, LFV rates strongly depend on the mass of the 
$3^{rd}$ generation heavy neutrino, so the rates are not expected to show a difference 
for a different set of neutrino mass choice with normal hierarchy since $M_{N_3}$ 
can only take values within a small range. 
With IH, it is always possible to suppress SUSY enhanced LFV observables by taking $M_{N_3} \sim m_{\rm GUT}$ as seen in Fig.~\ref{fig:evolml2}. 
We adopt $m_{\nu_3} \simeq 0.05$ eV as the common third generation neutrino mass. 
%which is at the higher end of the allowed range for the IH case.

\section{Lepton Flavor Violation and Observables}
\label{sec:flav}

In supersymmetry, non-diagonal mass matrix elements in the slepton mass matrix 
can be the source of LFV processes~\cite{Borzumati:1986qx}. 
For example, in the mass insertion (MI) method with leading $\log$ approximation, 
the branching fractions for processes $l_i \to l_j  \gamma$ can be approximated as:
\be
{\rm BF}(l_i \rightarrow l_j \gamma) \simeq 
\left|(m^2_{\widetilde L})(LL)_{ij}\right|^2 \frac{\alpha^3  \tan^2 \beta}{G_F^2 m_S^8}
\label{eq:brapp}
\ee
where $\alpha$ is the fine structure constant, $G_F$ is the Fermi constant, 
and $m_S$ defines the mass scale of the SUSY particles~\cite{Borzumati:1986qx}. 
In the MI technique, one defines the $6 \times 6$ slepton mass matrix as:
\be
m^2_{\widetilde L}  =  
 \begin{pmatrix}
m^2_{\widetilde L}(LL)_{ij} & m^2_{\widetilde L}(LR)_{ij}  \\
  &    \\
m^2_{\widetilde	 L}(RL)_{ij} & m^2_{\widetilde L}(RR)_{ij}\\
 \end{pmatrix}
\ee
where $LL$, $LR$, $RL$, and $RR$ are $3 \times 3$ entries defined based on the 
chirality label of the sfermions.
%Each term of the matrix is defined explicitly in Appendix A.\\

In the MSSM (with no RHNs), off-diagonal elements of the slepton mass matrix are 
small enough to not give rise to  significant flavor violating rates. 
Hence, flavor changing processes such as $\mu \to e\gamma $ are suppressed. 
When heavy right-handed neutrinos are introduced into the model, RG running 
can give rise to large values for the mass matrix elements. 
In $SO(10)$ SUSY models, the dominant contribution arising from the neutrino sector is 
proportional to the squared top-quark Yukawa coupling $Y_t^2$.
% For models that assumes a standard model like Yukawa coupling matrix and heavier %third generation neutrino, 
Hence, the size of the elements of the slepton mass matrix associated with 
LFV processes depends mostly on the running from $m_{\rm GUT}$ to $M_{N_3}$. 

\subsection{$\mu \to e  \gamma$}
\label{ssec:mu2e}

\begin{figure}[t]
  \centering
  {\includegraphics[width=.48\textwidth]{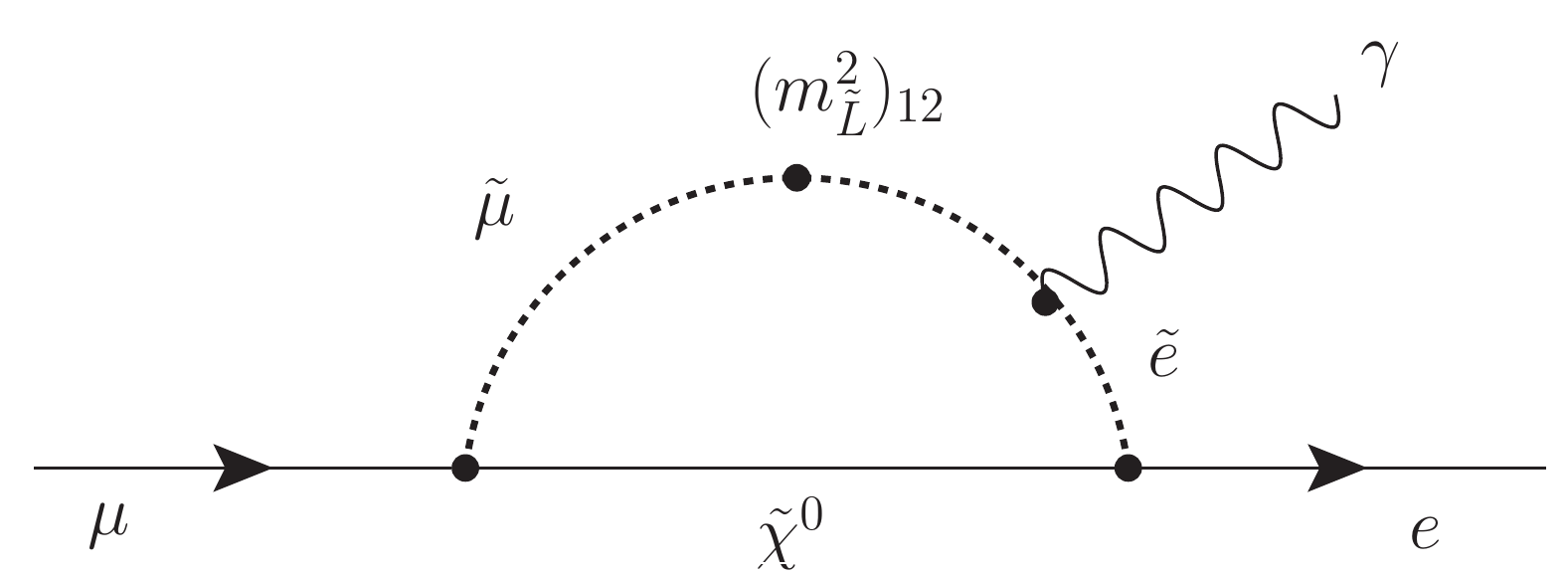}}\quad
  {\includegraphics[width=.48\textwidth]{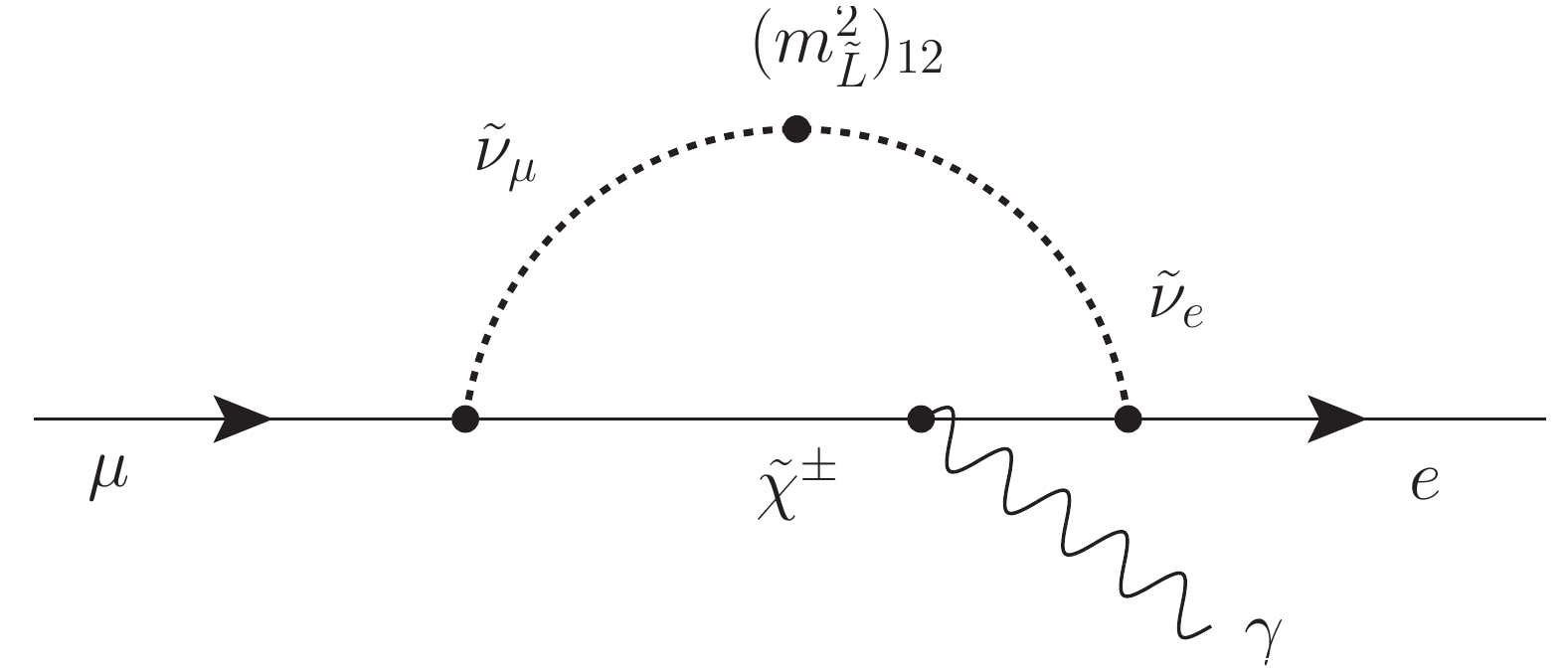}}\\
%\caption{A Feynman diagram contributing to $\mu\to e\gamma$ in MSSM+RHN models.}
\caption{Examples of Feynman diagrams contributing to $\mu\to e\gamma$ in MSSM+RHN models.}
\label{fig:diag}
\end{figure}

In supersymmetry, flavor changing processes are induced by one-loop diagrams with the 
exchange of gauginos and sleptons (see Fig.~\ref{fig:diag} for neutralino and chargino diagrams).
The leading log approximation for the slepton mass matrix element that induces the 
process $\mu \to e  \gamma $ reads:
\be
 (m^2_{\widetilde L})_{12} \simeq -\frac{1}{8\pi^2}(2m_0^2+m_{H_u}^2+A_0^2)
 \sum_k ({\bf f_{\nu}^T})_{1k}({\bf f_{\nu}}^*)_{k2} \log \frac{m_{\rm GUT}}{M_{N_k}} .
 \label{eq:loopapp}
\ee
For low values of $m_{1/2}$, the approximation holds well but for values $m_{1/2} \simeq 1$ TeV, 
the branching fractions can be up to a factor of 10 different than estimates from 
RGE running~\cite{Petcov:2003zb}. 
If the hierarchy of neutrino couplings are similar to that of the up-type in the 
Standard Model, where the third generation dominates, then the largest contribution 
is from the $k=3$ term in the summation. 
BF$(\mu \to e  \gamma)$ can be calculated by substituting the $m^2_{\widetilde L}$ term in 
Eq.~(\ref{eq:loopapp}) into the expression in (\ref{eq:brapp}) for a good approximation 
for a light/pre-LHC SUSY mass spectrum.

In the mass insertion technique, the branching fraction for the process $\mu \to e  \gamma$ 
can be calculated using:
\be
{\rm BF}(\mu \to e  \gamma) = \frac{48 \pi^3 \alpha}{G_F^2}(|A_L^{\mu e}|^2 + |A_R^{\mu e}|^2)
\ee
where the amplitudes $A_L$'s and $A_R$'s are given in Ref.~\cite{Hisano}.
%$\alpha$ is the electromagnetic structure constant and $G_F$ is the Fermi constant. 
In 2016, the MEG collaboration reported the current bound for the process as 
BF$(\mu \to e  \gamma) < 4.2 \times 10^{-13}$ at 90\% CL~\cite{TheMEG:2016wtm}. 
The expected sensitivity of the MEG-II experiment is $6\times 10^{-14}$ for three years 
of data taking~\cite{Baldini:2018nnn}. 

\subsection{$\tau \to \mu  \gamma$}
\label{ssec:taumu}

In a similar way to the leading log approximation for $\mu \to e  \gamma $ decay, 
BF$(\tau \to \mu  \gamma)$ can be approximated as:
\be
 (m^2_{\widetilde L})_{32} \simeq -\frac{1}{8\pi^2}(2m_0^2+m_{H_u}^2+A_0^2)
 \sum_k ({\bf f_{\nu}^T})_{3k}({\bf f_{\nu}}^*)_{k2} \log \frac{m_{\rm GUT}}{M_{N_k}}
 \label{eq:loopapp2}
\ee
where the dominant contribution comes from the term with 
$({\bf f_{\nu}^T})_{33}({\bf f_{\nu}}^*)_{32} \propto Y_t^2 V_{tb} V_{ts}$ in models with 
CKM-like neutrino Yukawa mixings and from the term $\propto Y_t^2 U_{\mu 3} U_{\tau 3}$ 
in models with PMNS-like mixings~\cite{Calibbi:2012gr}. 
The most stringent current constraint come from BaBar collaboration 
BF$(\tau \to \mu  \gamma)<4.4 \times 10^{-8}$ at 90\% CL~\cite{Aubert:2009ag};
the Belle-II experiment expects to gain a factor of 10 improvement~\cite{BelleII}.

\subsection{$\mu \to 3e$}

In SUSY, the decay $\mu \to 3e$ gets contributions from $\gamma$-, $Z$-, $Higgs$-penguin and 
box diagrams where the $\gamma$-penguin is the dominant one~\cite{Arganda:2005ji}. 
The leading $\gamma$-penguin approximation gives the relation :
\be
\frac{{\rm BF}(\mu \to 3 e)}{{\rm BF}(\mu \to e \gamma)}=\frac{\alpha}{3 \pi}
 \begin{pmatrix}
\log \frac{m_{\mu}^2}{m_e^2} - \frac{11}{4} \\
 \end{pmatrix}
 .
\ee
The current experimental lower bound on BF$(\mu\to eee)$ is from the SINDRUM experiment : 
$10^{-12}$~\cite{Bellgardt:1987du}. 
The proposed Mu3e experiment has a projected sensitivity of 
$5.2 \times 10^{-15}$~\cite{Perrevoort:2018ttp} up to $\sim 10^{-16}$ at Phase II. 

\subsection{$\mu N \to e N$}
\label{ssec:mu2econ}

The $\mu\to e$ conversion rate (CR) inside a nucleus $N$ depends on the neutron and proton densities. 
The rate is larger with increasing $Z$ (atomic number) up to $Z\sim 30$ and decreasing 
for $Z \gtrsim 60$~\cite{Kitano:2002mt}. 
The Mu2e experiment is designed to look for $\mu N \to e N$ conversion with a 
sensitivity of conversion rate CR$(\mu + Al \to e + Al) \sim 2.4 \times 10^{-17}$ 
in aluminum~\cite{Bartoszek:2014mya}. 
The sensitivity is improved by replacing aluminum targets with titanium ones. 

\section{Results for SUSY models with non-universal scalar masses}
\label{sec:gl}

The cMSSM+RHN model parameter set has been well-investigated in the 
literature~\cite{Masiero,Hirsch:2012ti,Barger:2009gc}. 
Also, the NUHM1,2+RHN models were previously studied for specific non-universal 
scenarios with the following GUT-scale mass relations: 
NUHM1 ($m^2_{H_u}=m^2_{H_d}\neq m^2_0$) in Ref.~\cite{Calibbi:2012gr}, 
NUHM2 ($m^2_{H_u}\neq m^2_{H_d}\neq m^2_0$) and 
NUGM (non-universal gaugino masses with $M_1:M_2:M_3={−1/2}:−3/2:1$) in Ref.~\cite{Bora:2014mna}. 
In these papers, the authors fixed the RH neutrino masses at $m_{\rm GUT}$.
%which gives the observed light neutrino mass gaps at the weak scale. 
Here, we adopt the three-extra-parameter non-universal Higgs mass SUSY model NUHM3 
wherein the first and second generations of matter scalars 
are split from the third ($m^2_{H_u}\neq m^2_{H_d}\neq m^2_0(3)\neq m^2_0(1,2)$).
The NUHM3 model is well motivated in that it allows for natural SUSY with 
$\Delta_{\rm EW}<30$ (and the required small $\bar{\mu}$ parameter) 
while also allowing for third generation non-unversality as is expected
in many stringy constructs~\cite{Baer:2017uvn,Nilles:2014owa}.

We scan over the following parameter space:
\begin{table}[h!]
\begin{tabular}{llllll}
MSSM parameters:   & & & & & sneutrino parameters: \\

$m_0(3)$   &:  $0-20 \:\: {\rm TeV}$ & & & & $m_{\tilde{\nu}_3}=m_0(3)$  \\
$m_0(1,2)$   &:  $0-20 \:\: {\rm TeV}$ & & & & $m_{\tilde{\nu}_{1,2}}=m_0(1,2)$  \\
$m_{1/2}$   &:  $0-4 \:\: {\rm TeV}$ & & & & $A_{\nu}=A_0$\\
$A_0$ &: $(-4-4)\:\: m_0(3) $   &  & &  &  \\
$\bar{\mu}$ &: $0-1 \:\: {\rm TeV} $ & & &  &  \\
$m_A$ &: $0-10 \:\: {\rm TeV} $   &  & &  &  \\
$\tan \beta$ &: $3-60$   &  & &  &  $m_{\nu_3} \simeq 0.05$ eV (weak scale) .
\end{tabular}
\label{tab:scan}
\end{table}\\
The first four SUSY breaking masses are input at the GUT scale; 
$\bar{\mu}$ and $m_A$ are input at the weak scale. 
Weak scale values of $\bar{\mu}$ and $m_A$ are preferred instead of $m_{H_u}$ and $m_{H_d}$ at GUT scale 
for a better statistical sampling of $\bar{\mu} \leq$1 TeV. 
We also assume $\bar{\mu}$ to be positive and real. 
The lower-end limits of SUSY breaking terms and $\bar{\mu}$ are in tension with LHC and LEP2 searches.
More precisely, $m_{\tilde{g}}>2.25$ TeV and $m_{\tilde{\chi}_1^{\pm}}>103.5$ GeV require 
$m_{1/2}\gtrsim 1$ TeV and $\bar{\mu}\gtrsim 100$ GeV respectively.
We pay some special attention to {\it natural SUSY} solutions with $\Delta_{\rm EW}<30$ since these
generate a weak scale of $m_{weak}\sim 100$ GeV without requiring any large fine-tunings in 
Eq.~(\ref{eq:mzs}).

Only the solutions that give the neutrino mass spectrum of interest 
($m_{\nu_3}=0.05\pm 0.0025$ eV and $m_{\nu_{1,2}}$ determined by $\Delta m_{ij}^2$ 
given in Ref.~\cite{Capozzi:2017ipn}) after the RGE running are accepted. 
Results are shown in two different categories based on the {\it sneutrino} mass hierarchy 
at the weak scale:\\
\vspace{-0.2cm}

\noindent $m_{\tilde{\nu}_3} > m_{\tilde{\nu}_2} \simeq m_{\tilde{\nu}_1}$ $\to$ normal sneutrino hierarchy \:(NSH),\\
$m_{\tilde{\nu}_3} < m_{\tilde{\nu}_2} \simeq m_{\tilde{\nu}_1}$ $\to$ inverted sneutrino hierarchy (ISH).\\
\vspace{-0.2cm}

Since first and second generation masses can drastically increase during RGE running for 
heavy enough gluinos, imposing a hierarchy on scalars ($m_0(1,2,3)$'s) at the GUT scale 
does not guarantee that one generates the same hierarchy at the weak scale. 
Furthermore, LFV processes such as $\mu \to e \gamma$ and $\mu \to e e e$ are directly 
related to the sneutrino mass hierarchy rather than the scalar mass hierarchy at the weak scale. 
In the rest of the paper, we use the terms NSH and ISH for the weak scale hierarchy 
of sneutrinos.

We adopt a top-down approach: all the parameters are evolved from $m_{\rm GUT}$ to $m_{weak}$ with 
the RH neutrinos decoupled one-by-one using the methodology of
Antusch {\it et al.}~\cite{Antusch:2005gp}. 
SUSY spectra are calculated by using ISAJET~\cite{Paige:2003mg} with neutrino related RGEs 
calculated using a modified version of the subroutine ISABRS that comes within the ISAJET package. 
After RG running from $m_{\rm GUT}$ to $m_{weak}$, the mass matrices $m_{\tilde{L}}^2$, $m_{\tilde{\nu}}^2$, 
$m_{\tilde{\chi}^0}^2$ and $m_{\tilde{\chi}^\pm}^2$ are calculated at the weak scale. 
We accept solutions that give the observed neutrino oscillations and with a dark matter abundance 
less than the measured result, $\Omega h^2 \leq 0.12 $ 
(assuming that the remainder of the dark matter might be composed of axions). 
Light neutrino masses are calculated by diagonalizing the coupling of the dim-5 operator, $\kappa$. 
We use the package SUSEFLAV~\cite{Chowdhury:2011zr} to calculate BF$(\mu \to e \gamma)$, 
BF$(\tau \to \mu \gamma)$ and BF$(\mu \to eee)$ by feeding in the low energy spectrum 
after RG running of ISAJET. 
SUSEFLAV calculates flavor changing observables using MIs. 
To calculate the CR$(\mu N \to e  N)$ for aluminum and titanium nuclei, 
we adopt the values for the effective nuclear charge $Z_{eff}^{Al}=11.6$ and the form factor 
$F(q)=0.64$ for $^{27}_{13}Al$ and  $Z_{eff}^{Ti} \simeq 17.6$ with $F(q)=0.52$ for 
$^{48}_{22}Ti$~\cite{Kitano:2002mt}.

The main contribution to off-diagonal elements of the slepton mass matrix arises during 
the evolution from $Q=m_{\rm GUT} \to M_{N_3}$-- since the BF$(\mu \to e  \gamma) \propto  (f_{\nu})_{3,3}^2$-- 
so the mass of the third generation heavy neutrino plays a crucial role for the LFV processes. 
Our choice of $M_{N_3}$ generates nearly maximal LFV effects for both sneutrino hierarchies
NSH and ISN. 

We only show solutions with :
\begin{itemize}
\item neutralino LSP,
\item radiatively broken electroweak symmetry,
\item the lightest chargino mass greater than 103.5 GeV (LEP2 limit)~\cite{LEP2},
\item Higgs mass $m_h= 125 \pm 2$ GeV,
\item $1.12 \times 10^{-9} < {\rm BF}(B_S \to \mu^+ \mu^- ) < 4.48 \times 10^{-9}$~\cite{Endo:2015oia},
\item $2.79 \times 10^{-4} < {\rm BF}(b \to s\gamma) < 4.63 \times 10^{-4}$~\cite{Endo:2015oia}.
\end{itemize}

\begin{figure}[t]
  \centering
  {\includegraphics[width=.48\textwidth]{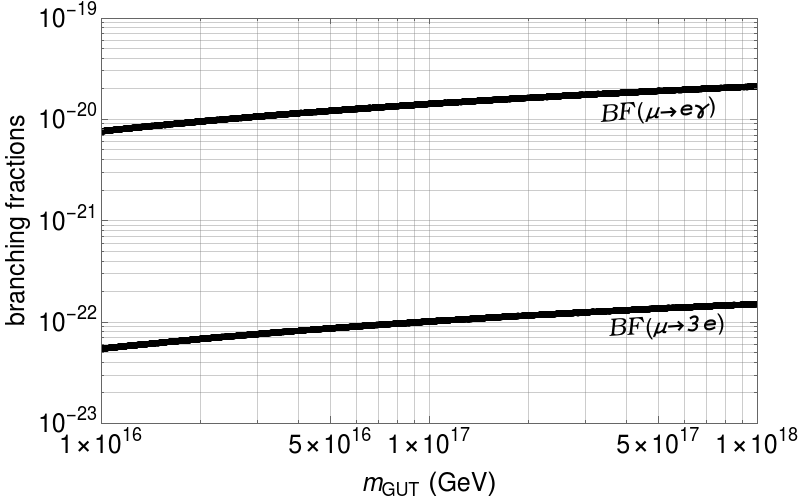}}\quad
  {\includegraphics[width=.48\textwidth]{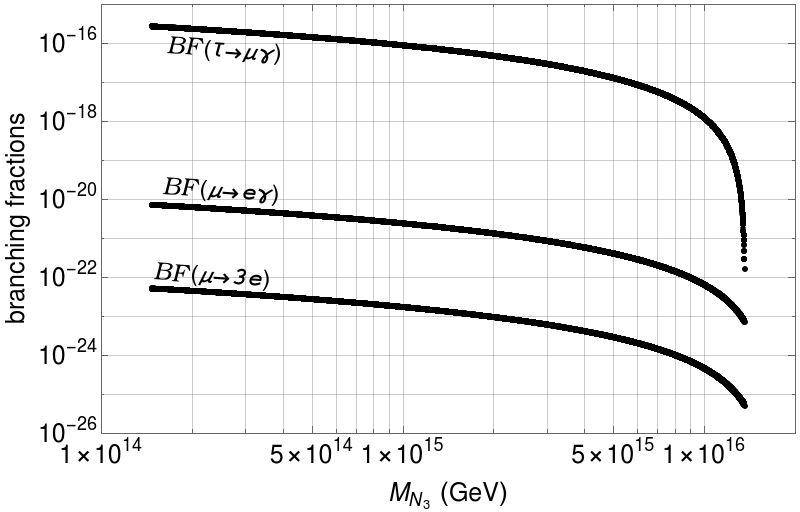}}\\
\caption{Dependence of BF($\mu \to e  \gamma$) (left) to $m_{\rm GUT}$ with fixed $M_{N_{1,2,3}} \equiv 10^{6,10,14}$ GeV and (right) to $M_{N_3}$ for inverse hierarchical light neutrinos.}  
\label{fig:d2}
\end{figure}

In Fig.~\ref{fig:d2}(left), we show the effect of taking $m_{\rm GUT}$ larger than the 
SUSY unification scale on the BF$(\mu \to e  \gamma)$ while keeping $M_{N_3}$ fixed at $10^{14}$ GeV. 
It is possible to assume $m_{\rm GUT}$ close to $M_{\rm string}$ which is considered to be 
slightly higher than the gauge coupling unification scale due to string loop 
effects~\cite{Witten:1996mz}. 
Taking $m_{\rm GUT}$ larger than the gauge coupling unification scale increases the rates for LFV 
processes  since $M_{N_3}$ is fixed. 
The effect of increasing $m_{\rm GUT}$ by {\it two} orders of magnitude increases LFV effects 
by a factor of {\it two} which is what we expect from the leading log approximation, 
Eq.~(\ref{eq:loopapp}). 
In the remainder of our analysis, we take $m_{\rm GUT}$ to be equal to the gauge coupling unification scale: 
$\sim 2.4 \times 10^{16}$ GeV.

In Fig.~\ref{fig:d2}(right), we show how sensitive LFV processes are to our choice of 
neutrino mass spectrum in the IH case. 
In the IH case, $M_{N_3}$ can take values from $\sim 10^{14}$ GeV 
(from cosmological bound on sum of neutrino masses) 
to $\sim m_{\rm GUT}$ $(0.001 < m_{\nu_3}/{\rm eV} < 0.1)$ whereas $1^{st}$ and $2^{nd}$ generations 
are more constrained. 
As seen in the figure, the branching fractions drop significantly for 
$M_{N_3} \gtrsim 5 \times 10^{15}$ GeV (corresponding to $m_{\nu_3}<0.002$ eV) since 
SUSY contributions to LFV approach $zero$ for $M_{N_3} \sim m_{\rm GUT}$. 
Hence, it is hard to rule out models with the IH unless there is a lower bound on 
$m_{\nu_3}$ from observations.

\subsection{Results for branching fractions}
\label{ssec:BRs}

\begin{figure}[t]
  \centering
  {\includegraphics[width=.48\textwidth]{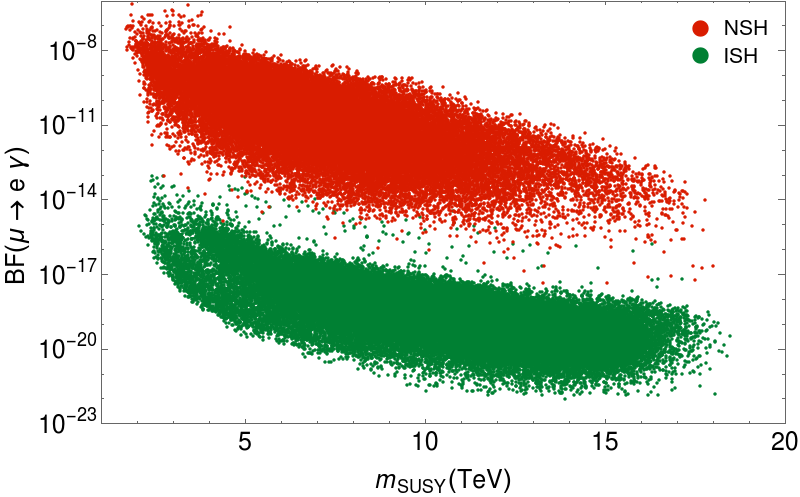}}\quad
  {\includegraphics[width=.48\textwidth]{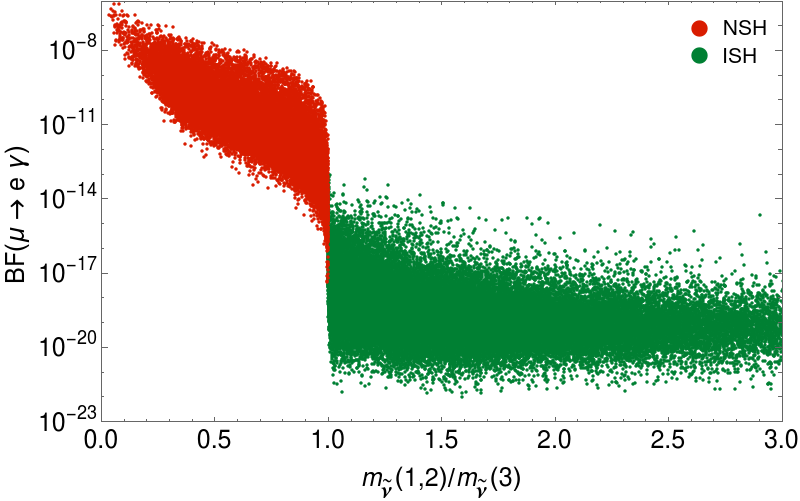}}\\
\caption{BF($\mu \to e  \gamma$) vs. $m_{\rm SUSY}$ (left) and vs. ratio of the $1^{st}/2^{nd}$ to 
$3^{rd}$ generation sneutrino mass $m_{\tnu_{1,2}}/m_{\tnu_3}$ (right), 
both with CKM-like neutrino mixing. Red: NSH and green: ISH.}  
\label{fig:corr}
\end{figure}

The branching fractions for leptonic decays $l_i \to l_j  \gamma$ are inversely proportional to 
$m_{\rm SUSY}^8$ in the leading log approximation where $m_{\rm SUSY}$ is the 
sparticle mass scale~\cite{Petcov:2003zb}. 
The correlation between $m_{\rm SUSY}$ and BF$(\mu \to e  \gamma)$ 
is shown in Fig.~\ref{fig:corr}(left). The BF$(\mu \to e  \gamma)$ is decreasing with increasing 
$m_{\rm SUSY}$ as expected~\cite{Borzumati:1986qx,Arana-Catania:2013xma}. 
Here we take $m_{\rm SUSY}$ as the arithmetic mean of the sfermion masses and the gluino mass. 
We constrain the parameter space to be within the HE-LHC sparticle mass reach~\cite{Baer:2018hpb}: 
$m_{\tilde{g}} \lesssim 6$ TeV.
The main feature of the NUHM3 model investigated here arises from the ratio of
$1^{st}/2^{nd}$ to $3^{rd}$ generation sneutrino masses as shown in Fig.~\ref{fig:corr}(right). 
Due to cancellations and decouplings within the amplitudes of chargino and neutralino loop diagrams, 
LFV observables are highly suppressed for sparticles with the ISH. 
Hence, $\mu \to e  \gamma$ decays are highly suppressed for $m(3) < m(2) \simeq m(1)$. 
This behavior is expanded upon in Appendix~\ref{sec:lc}.

\begin{figure}[h!]
  \centering
  {\includegraphics[width=.48\textwidth]{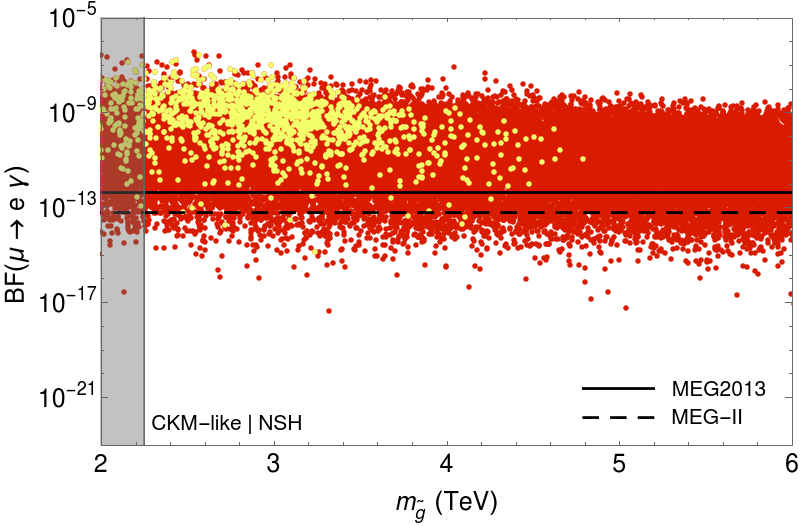}}\quad
  {\includegraphics[width=.48\textwidth]{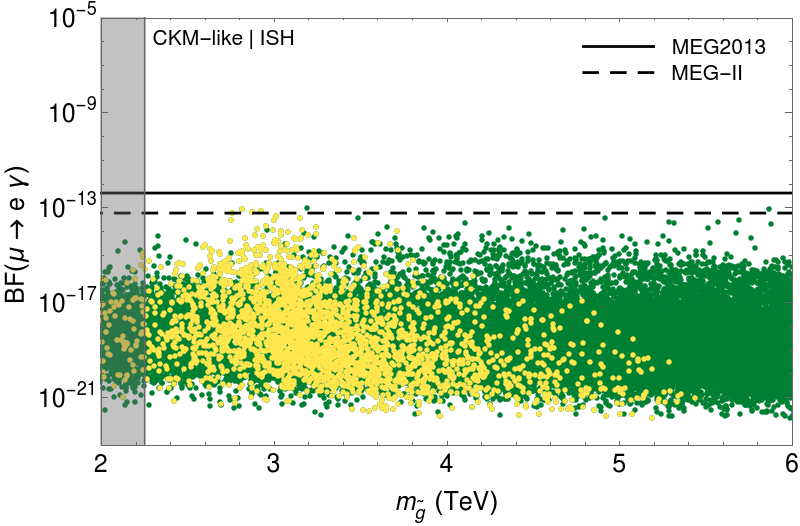}}\\ 
  {\includegraphics[width=.48\textwidth]{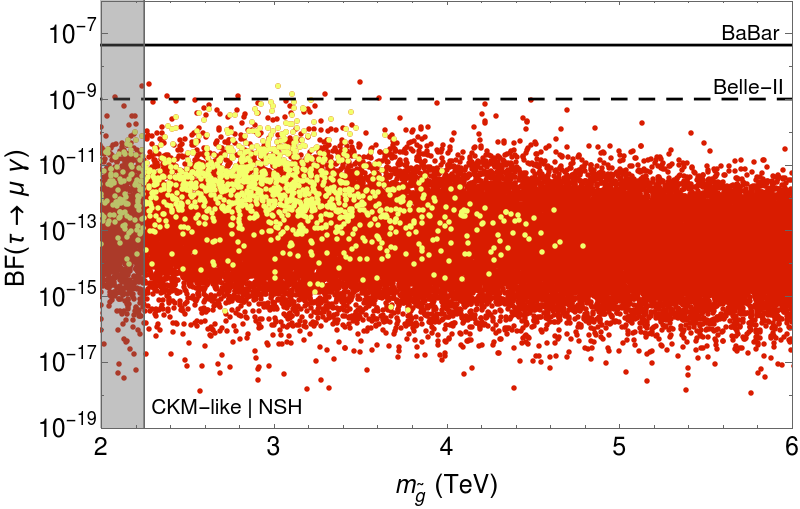}}\quad
  {\includegraphics[width=.48\textwidth]{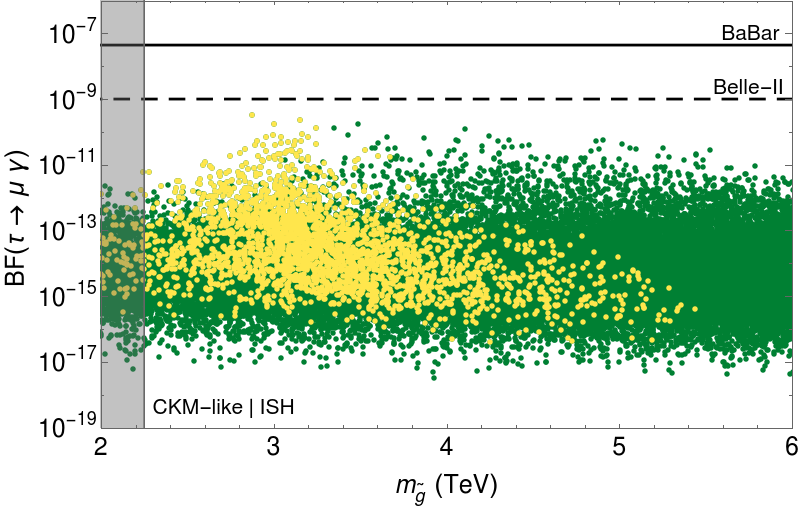}}\\ 
  {\includegraphics[width=.48\textwidth]{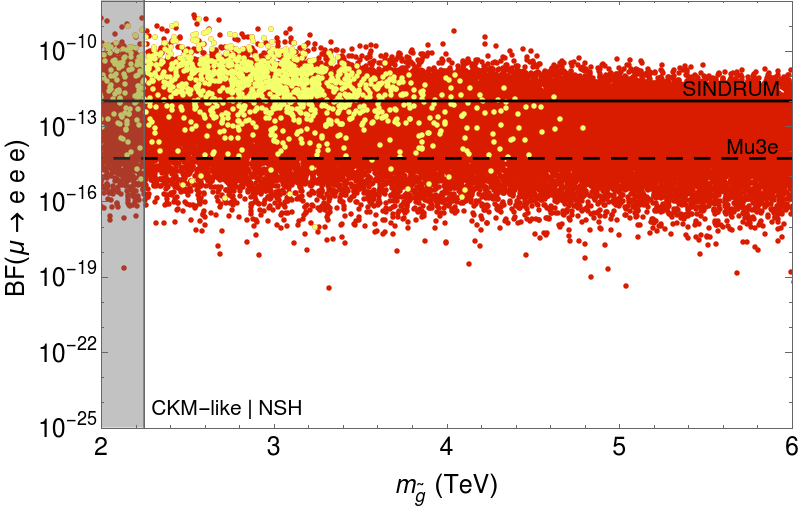}}\quad
  {\includegraphics[width=.48\textwidth]{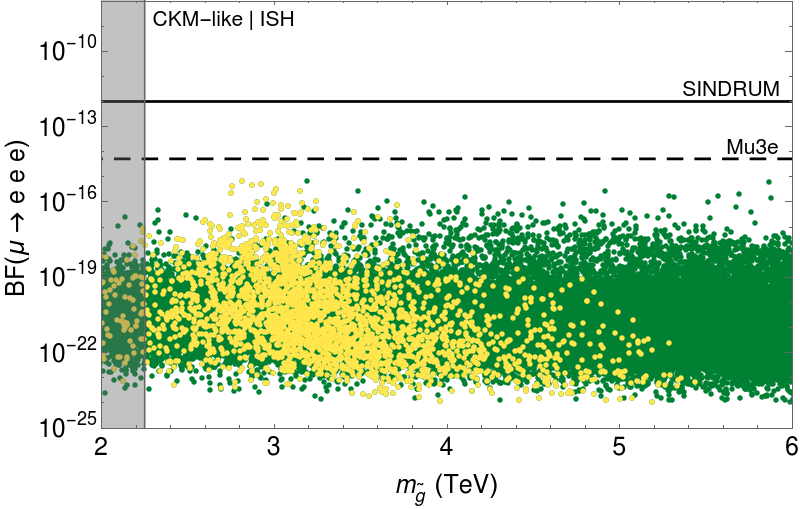}}\\ 
\caption{LFV observables for NSH (left) and ISH (right) for the Case \#1 small (CKM-like) mixing scenario. 
Red points denote NSH while green points denote ISH. 
Yellow points are natural with $\Delta_{\rm EW}<30$. 
The gray shaded region is excluded by LHC Run 2 gluino search constraints.}  
\label{fig:yuk1}
\end{figure}

\subsubsection{Scenario \#1: small CKM-like mixing, $({\bf f}_\nu)_{ij}=({\bf f}_u)_{ij}$}

Our results for LFV observables are shown in Fig.~\ref{fig:yuk1} where we show branching 
fractions for the  LFV decays $\mu \to e  \gamma$, $\tau \to \mu  \gamma$ and 
$\mu \to eee$ as a function of $m_{\tg}$ for sneutrino masses with the NSH (left) and with 
the ISH (right).
Current limits are shown by the solid horizontal lines while projected reaches of 
future LFV experiments MEG-II, Belle-II and Mu3e are shown by dashed lines.

For the small (CKM-like) mixing case, 
the transition from normal to inverted sneutrino mass hierarchy 
shows a sharp decrease for the branching fractions $\mu \to e  \gamma$ and $\mu \to eee$ 
when the dominant contribution to the amplitude $A_{L,R}^{\mu e,\mu eee}$ from the chargino loop 
$A_{L,R}^{C}$ changes sign which strongly depends on the hierarchy of sneutrino masses. 
The transition is smoother in the large mixing (PMNS-like) case since the cancellations between 
$A_{L,R}^{i,j}$ are not as strong as in the small mixing case.

From Fig.~\ref{fig:yuk1}, we see that the MEG2013 result has ruled out the NSH (left) of 
sfermions for the majority of the parameter space. 
This includes the bulk of natural SUSY points (yellow).
Surviving points have highly degenerate mass spectra $m(3) \simeq m(2) \simeq m(1)$ 
which might be expected from NUHM2 boundary conditions. 
Constraints from the other search channels $\tau \to \mu \gamma$ and $\mu \to eee$ are less stringent. 

The ISH case (right) in all channels predicts very mild flavor 
violation which may not be within the reach of projected sensitivities. 
The ISH with $m_0(3)\ll m_0(1,2)$ is actually favored by SUSY string landscape 
predictions which prefer soft SUSY
breaking masses as large as possible subject to appropriate EWSB~\cite{Baer:2017uvn}. 
In this case, all natural SUSY points lie below projected limits. 
Although BF$(\tau \to e  \gamma)$ is large due to large CKM mixing element 
$|V_{tb}| \sim \mathcal{O}(1)$, the Belle-II sensitivity is still higher than the 
predicted $\tau \to e  \gamma$ decay rates.

The predictions for natural SUSY (yellow points) with a NSH (left) are all above the existing 
MEG2013 limits. 
This is mainly because from naturalness $m_0(3)$ cannot exceed $\sim 5-10$ TeV lest 
contributions to the weak scale become large and the model becomes unnatural. 
For the NSH, the first/second generation sneutrino masses are necessarily lighter than third,
so this acts to enhance the LFV rates.
%In our scan, the upper limits for the scalar masses are 20 TeV (at GUT scale) 
%so the branching fractions are further suppressed by the heavier spectrum. 

For the (string preferred) ISH scenario (Fig.~\ref{fig:yuk1}(right) yellow points), 
natural SUSY predicts much lower rates for LFV branching fractions because now  
first/second generation sleptons can range up to the scan upper limits (as large as 20 TeV) 
with little effect on naturalness whilst third generations scalars, which are more strongly related to naturalness, remain in the several TeV range. 
Hence, points with $m_0(1,2)\simeq 20$ TeV (upper scan limit) set the lower bound for LFV processes 
for ISH for natural SUSY. 
As the gluino mass gets larger, the yellow points tend to populate lower BF values 
mainly because naturalness can be preserved with smaller $\tan \beta$, $\tan \beta < 20$ 
as $m_{1/2}$ gets larger.
The overall conclusion from the ISH plots of Fig.~\ref{fig:yuk1} is that the 
stringy preferred~\cite{Nilles:2014owa,Baer:2017uvn} natural SUSY models tend 
to predict LFV decay rates well below present and even future
experimental search limits.

\subsubsection{Scenario \#2: large PMNS-like mixing, $({\bf f}_\nu)_{ij}=({\bf f}_u)_{ii}^{\rm diag} \: {\rm \bf U}^{\rm PMNS}_{ij}$}

In this Subsection, we present predictions for LFV decays from the NUHM3 model but with 
Scenario \#2 large PMNS-like mixing of neutrino couplings. 
The cMSSM with PMNS-like mixing is already under tension from MEG results whereas the 
NUHM1 model with $\tan \beta < 20$ survives~\cite{Calibbi:2012gr,Bora:2014mna}. 

\begin{figure}[h!]
  \centering
  {\includegraphics[width=.48\textwidth]{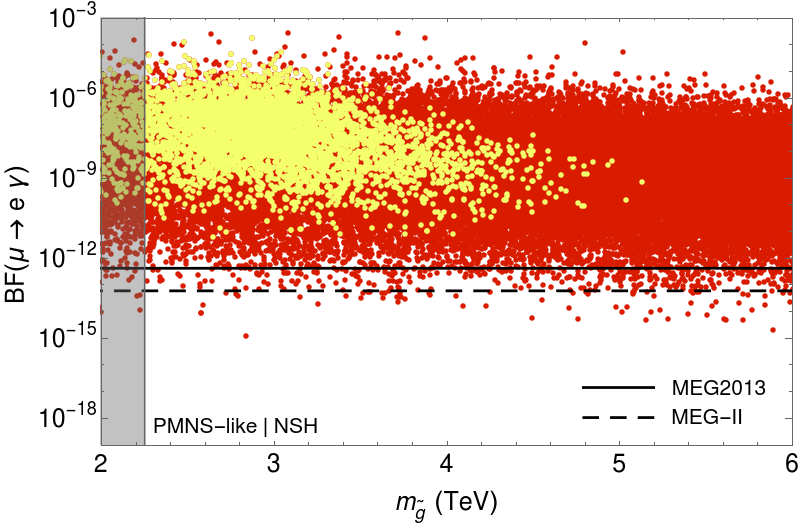}}\quad
  {\includegraphics[width=.48\textwidth]{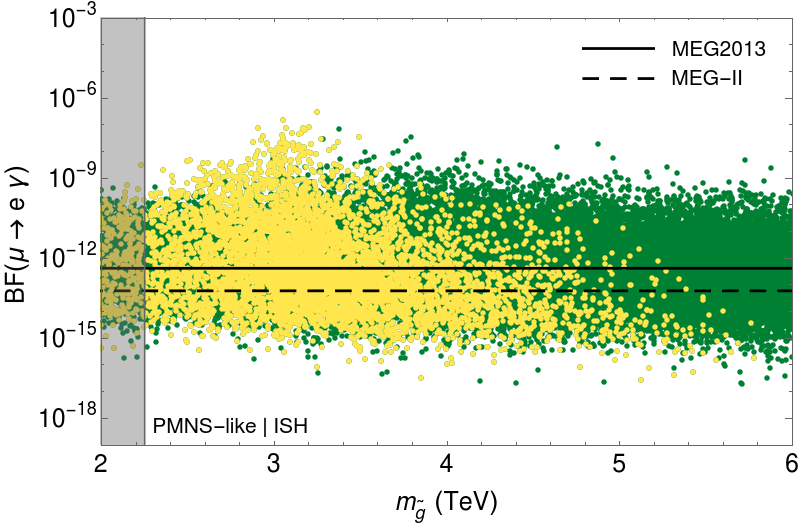}}\\ 
  {\includegraphics[width=.48\textwidth]{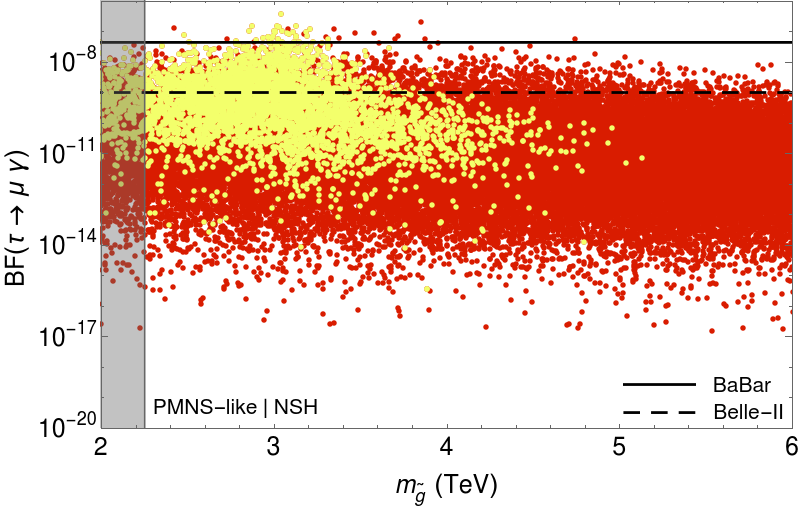}}\quad
  {\includegraphics[width=.48\textwidth]{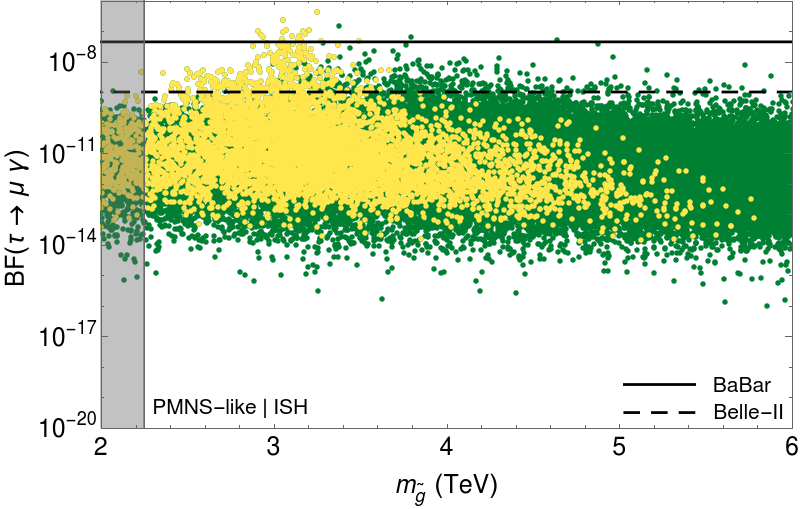}}\\ 
  {\includegraphics[width=.48\textwidth]{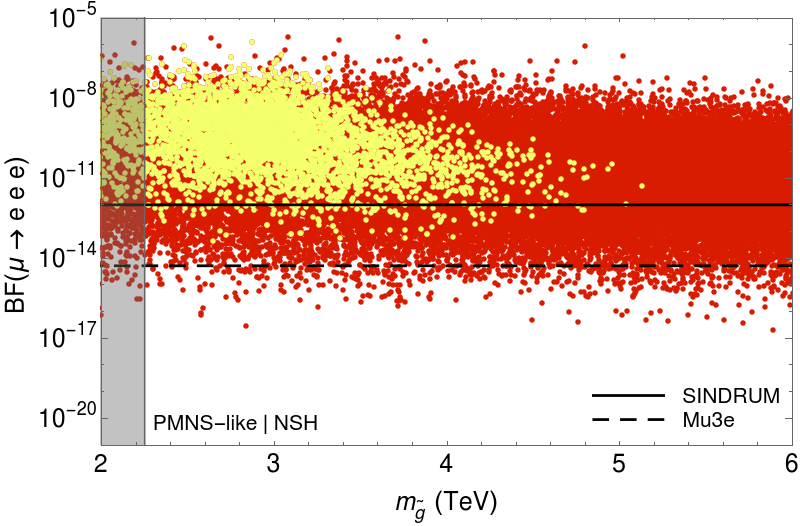}}\quad
  {\includegraphics[width=.48\textwidth]{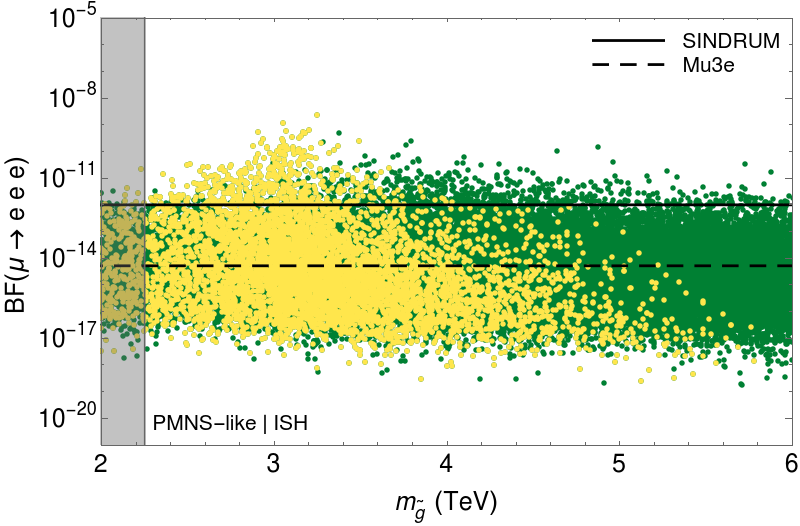}}\\ 
\caption{LFV observables for NSH (left) and ISH (right) for the Case \#2 large (PMNS-like) mixing scenario. 
Red points denote NSH while green points denote ISH. 
Yellow points are natural with $\Delta_{\rm EW}<30$. 
The gray shaded region is excluded by LHC Run 2 gluino search constraints.}  
\label{fig:mns}
\end{figure}

Predictions from the NUHM3 model for the LFV branching fractions BF$(\mu \to e  \gamma)$, 
BF$(\tau \to \mu  \gamma)$ and BF$(\mu \to eee)$ are shown in Fig.~\ref{fig:mns}
for the large PMNS-like mixing case for the NSH (left) and ISH (right). 
The large PMNS-like mixing predicts higher LFV rates so more of the parameter space 
is within the reach of future experiments compared to that of small CKM-like mixing. 
Similar to the CKM-mixing case, models with a NSH generate a few orders of magnitude 
larger rates compared to the ISH scenario.

For the large PMNS-like mixing scenario, the NSH is nearly {\it completely ruled out} by the MEG2013 results. The surviving points feature a heavy spectra for SUSY particles, $m_{\rm SUSY}\agt 10$ TeV.
For a heavier neutrino spectrum, LFV rates will be even more enhanced 
so our conclusions for the NSH case are valid for any set of neutrino masses with normal hierarchy ($m_{\nu_3} \gtrsim 0.05$ eV).
%In NUHM3, both $m_{H_u}$ and $m_{H_d}$ are free parameters and hence flavor violating operators do not show a strong correlation with the value of $\tan\beta$ unlike the NUHM1 model~\cite{Calibbi:2012gr}. 
Unlike the NUHM1 model results of Ref.~\cite{Calibbi:2012gr}, $\tan \beta$ values up to $\sim 50$ 
are still allowed for the NUHM3 model. 
%MEG2013 rules out the parmeter space with $\tan \beta > 50$ for $m_{\tilde{g}} \simeq 2$ TeV. 
%If no signal will be observed at MEG-II,  $\tan \beta > 50$ will be excluded for the 
%whole gluino mass range.{\bf Since we do not show any tanbeta dependence, shall we include this discussion?}

For the ISH case with large PMNS-like mixing, the MEG2013 limit rules out about half 
of the scanned parameter space points. 
The MEG-II experiment can probe more of-- but not all of-- 
the large mixing NUHM3 parameter space with an ISH.

Natural SUSY points show a similar behavior as in small CKM-like mixing case: 
smaller LFV rates for all decays in the ISH case as compared with the NSH. 
MEG-II and Mu3e searches can probe much of the NUHM3 large mixing ISH
parameter space but are not sensitive enough to cover its entirety. 
Points over the entire range of gluino masses can escape from all search channels 
since in NUHM3 $\tan \beta$ values less than 20 are allowed and these predict small LFV rates.

\subsection{$\mu \to e $ conversion in nuclei}

\begin{figure}[tbp]
  \centering
  {\includegraphics[width=.48\textwidth]{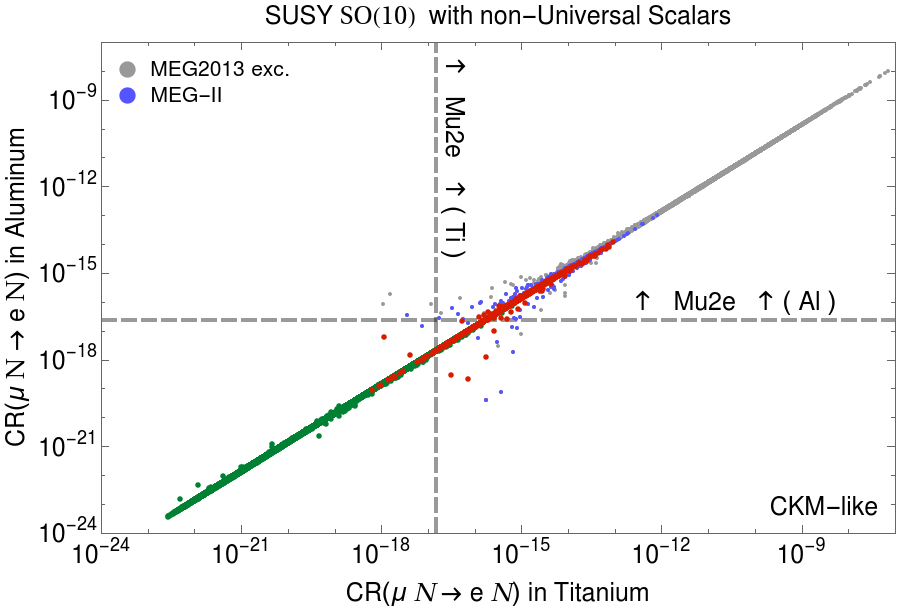}}\quad
  {\includegraphics[width=.48\textwidth]{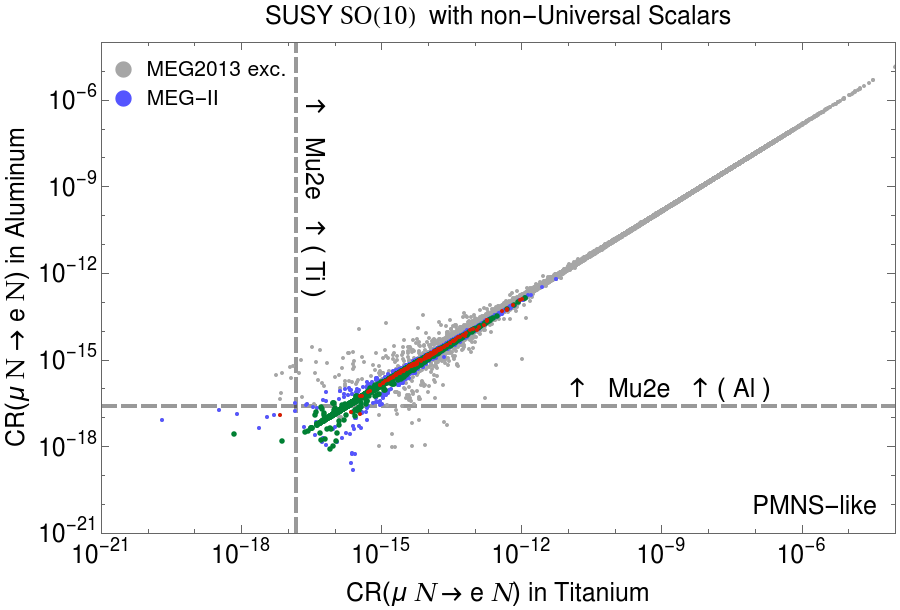}}\\
\caption{Conversion rate (CR) for $\mu N \to eN$ in aluminum and in titanium nuclei 
with  small CKM-like mixing (left) and large PMNS-like mixing (right). 
MEG2013 excluded points and points within MEG-II reach are colored in gray and blue respectively. 
The dashed gray lines show the projected sensitivity of the Mu2e experiment.
Red points denote the NSH while green points denote the ISH.}  
\label{fig:altipmns}
\end{figure}

In Fig.~\ref{fig:altipmns}, we focus on $\mu N \to e N$ conversion in aluminum and titanium nuclei. 
The Mu2e experiment is planned to initially start with aluminum targets which can be replaced 
with titanium ones without any major changes in detectors~\cite{Bernstein:2019fyh}. 
Titanium targets are expected to increase the sensitivity by a factor of 
$\sim$1.6~\cite{Kitano:2002mt,Cirigliano:2009bz}. 
%The projected Mu2e reaches in $Al$ and $Ti$ nuclei are shown by the dashed horizontal and vertical 
%lines, respectively.

The projected initial sensitivity of Mu2e -- CR$(\mu + Al \to e + Al) \sim 2.4 \times 10^{-17}$-- 
is shown by the horizontal dashed line. 
The approximate sensitivity with the titanium target is shown by the vertical dashed line. 
We also mark the points that are within the reach of MEG-II searches (blue) 
and points that are already ruled out by MEG2013 (gray) on the same plane. 
Only the points which are currently safe from LHC searches are shown in 
Fig.~\ref{fig:altipmns}.

The $\mu N \to e N$ conversion rates for small CKM-like mixing are shown on the left side of 
Fig.~\ref{fig:altipmns}. 
With the titanium target upgrade, the Mu2e experiment will probe almost the entire parameter space 
with NSH (red) but only a small portion of the parameter space predicted by models with 
the ISH (green).
For the large PMNS-like mixing scenario (right-side), even the initial run of Mu2e will
cover all of the NSH parameter space and a large portion of the ISH parameter space. 
The Mu2e will reach sufficient sensitivity to cover 
the entirety of the large PMNS-like mixing parameter space for both the NSH
and ISH cases with the titanium target upgrade. The proposed Mu2e-II upgrade envisions to improve the sensitivity by an order of magnitude, CR$(\mu + Al \to e + Al) \sim 2.5 \times 10^{-18}$~\cite{Abusalma:2018xem} 
which is still not quite sufficient to cover the whole CKM-like mixing scenario with ISH.

\section{Summary and conclusions}
\label{sec:conclude}

The see-saw mechanism is highly motivated in that a large, intermediate mass scale for 
Majorana neutrinos explains the tiny active neutrino masses whilst also explaining the absence
of right-handed neutrino effects in low energy data. The see-saw mechanism isn't a particularly
well-motivated extension of the SM all by itself since the (apparently fundamental) newly discovered
Higgs boson mass would likely blow up to the see-saw scale due to its quadratic divergences.
Supersymmetry stabilizes the Higgs mass so the weak scale can co-exist along with the 
Majorana mass scale (and the GUT and Planck scales). 
Within models containing a SUSY see-saw mechanism,
then LFV processes should occur, possibly at an observable level. Our goal in this paper
has been to present predictions for LFV processes within plausible SUSY models that are compatible with LHC Run 2 results. 
These should include {\it natural SUSY} models with radiatively-driven naturalness~\cite{Baer:2012up}
which allow for a 125 GeV Higgs mass along with multi-TeV soft terms (as implied by LHC data)
while at the same time avoiding the fine-tunings associated with a Little Hierarchy:
why is $m_{weak}\ll m_{soft}$?
\vspace{0.3cm}

We have investigated the current and projected reaches of various LFV search
experiments within both the natural and unnatural portions of the NUHM3 SUSY model 
where neutrinos are generated with a type-I seesaw mechanism. We have limited the SUSY parameter space within the HE-LHC gluino reach.
The NUHM3 model is well motivated in that it allows for weak scale naturalness along with a
125 GeV Higgs mass and sparticles beyond LHC Run 2 limits. The generational non-degeneracy
is well motivated by both landscape~\cite{Baer:2017uvn} and mini-landscape~\cite{Nilles:2014owa} 
string-motivated models which tend to give $m_0(3)\sim$ few TeV whilst $m_0(1,2)\sim$ tens of TeV.
We evaluated $two$ scenarios for the neutrino couplings:
Scenario \#1 small CKM-like mixing with ${\bf f}_{\nu}={\bf f}_u$ and 
Scenario \#2 large PMNS-like mixing with ${\bf f}_{\nu}={\bf f}_u^{\rm diag}$ U$^{\rm PMNS}$. 
These scenarios are originally motivated by 4-d $SO(10)$ SUSY GUT models.
We should remark here that the large mixing case arises from 16s of $SO(10)$ which couple
to 120-dimensional Higgs representations. Such large Higgs representations do not appear in stringy
constructions~\cite{Halverson:2018xge} and hence may well occupy the swampland~\cite{Vafa:2005ui}:
theories inconsistent with string theory. Both the small and large mixing cases 
should be consistent with local GUT theories which can emerge from string theory~\cite{Buchmuller:2005sh}.
Indeed, heterotic string models compactified on certain orbifolds can easily include the
type-I SUSY see-saw while allowing for split multiplets which solve the doublet-triplet splitting
problem~\cite{Buchmuller:2007zd}.

In our analysis, the $third$ generation neutrino mass is fixed to 0.05 eV and then 
first and second generation neutrino masses are calculated using observed squared mass differences.
Along with large and small Yukawa mixing cases, we divided our results according to
normal and inverted {\it sneutrino} mass hierarchies, NSH and ISH.
Our calculations show that for natural SUSY with the NSH, rates for
BF$(\mu \to e \gamma)$ are {\it already ruled out} by the MEG2013 experiment
for both small and large mixing scenarios. 
The $\mu \to e \gamma$ decay rate for the (landscape-favored) ISH scenario is lower 
than the NSH case due to the loop cancellations and decoupling effects. 
Even so, large chunks of parameter space for the ISH with large mixing are already ruled out, 
although substantial regions remain viable.
In contrast, the LFV decay rates for the (string-favored) ISH with small CKM-like mixing 
have not yet been touched by LFV search experiments. 
Moreover, LFV predictions from these latter models lie below even the future attainable limits 
so it appears hard to be optimistic that meaningful constraints will be gained from LFV probes
in the most plausible SUSY scenarios.
In addition, the Mu2e conversion search experiment apparently can test (nearly) the entire
parameter space of both NSH and ISH models with large mixing. 
The small mixing case for the ISH model remains with predicted detection rates well below 
the projected sensitivity of $\mu\to e$ conversion experiments.
%Alternatively, for models with large PMNS-like mixing, 
%then both Mu2e and Mu3e experiments have been or will be probing  considerable portions 
%of the parameter space.

%Searches for $\mu \to e \gamma$ and $\mu \to e e e$ can be used to distinguish between 
%CKM- and PMNS-like mixings scenarios. 
%An observation of a $\mu \to e \gamma$ event by MEG-II or by Mu2e would point to two possible choices:% 
%CKM-like mixing with NSH or PMNS-like mixing with ISH. 
%An additional signal from the $\mu \to e e e$ search channel would favor PMNS-like mixing 
%since CKM-like mixing predicts much lower rates.

\section*{Acknowledgments}

This work was supported in part by the US Department of Energy, Office
of High Energy Physics.
The computing for this project was performed at the OU Supercomputing Center for Education \& Research (OSCER) at the University of Oklahoma (OU). The Feynman diagrams in Fig~\ref{fig:diag} were drawn with JaxoDraw~\cite{Binosi:2008ig}.

\newpage
\section*{Appendix}

\appendix

\section{Loop cancellations}
\label{sec:lc}

\begin{figure}[tbp]
  \centering
  {\includegraphics[width=.524\textwidth]{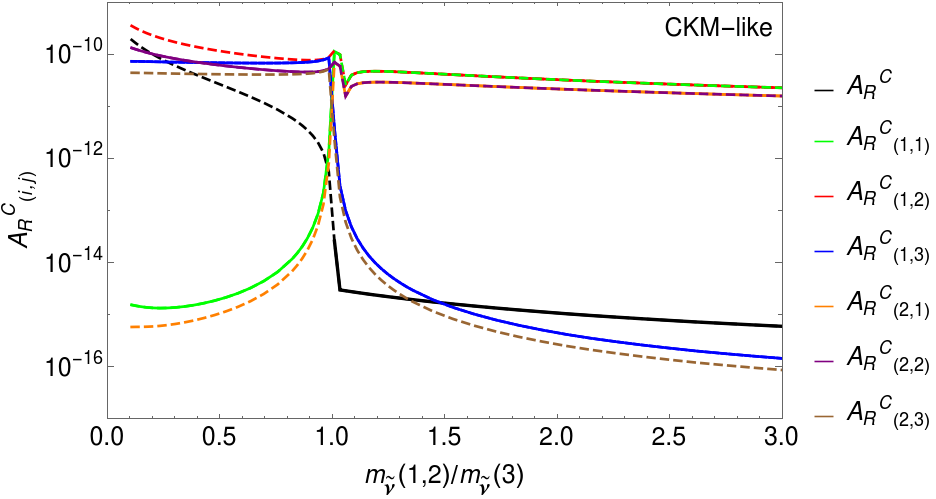}}\quad
  {\includegraphics[width=.45\textwidth]{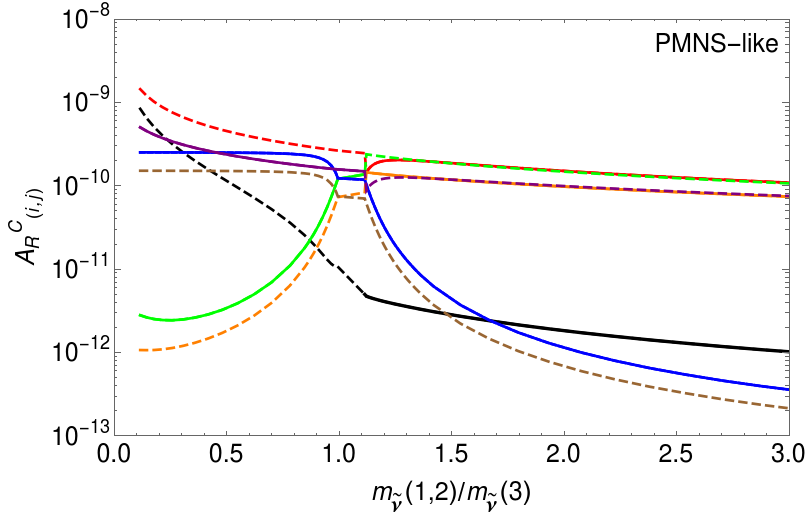}}\\ 
  \caption{Individual contributions to $A_R^C$ versus $m_{\tnu}(1,2)/m_{\tnu}(3)$. 
Dashed (solid) curves show negative (positive) amplitudes.}  
\label{fig:amp}
\end{figure}

The decay width for the process $\mu \to e  \gamma$ is given by:
\begin{equation}
\Gamma (\mu \to e  \gamma) =\frac{e^2}{16 \pi}m_{\mu}^5 (|A_L|^2+|A_R|^2)
\end{equation}
where $A_L=A_L^N+A_L^C$ and $A_R=A_R^N+A_R^C$ are the sum of the contributions from chargino ($C$) and neutralino ($N$)
loop diagrams. 
The main contribution to BF$(\mu \to e  \gamma)$ comes from the chargino sector. 
Amplitudes of contributions to $A_R^C$ are shown in Fig.~\ref{fig:amp} with respect to the 
sneutrino mass hierarchy for both small CKM-like (left) and large PMNS-like (right) mixing cases. 
The individual contribution $A^C_{R(i,j)}$ where $i$ and $j$ are the indices  for the chargino ($\chi^{\pm}_{1,2}$) 
and sneutrino ($\tilde\nu_{1,2,3}$) generations respectively, are defined in Ref.~\cite{Hisano}.
Dashed lines represent negative amplitudes whereas solid lines are positive amplitudes. 
The black line shows the sum of all the individual contributions: $A_R^C$. 
For $m_{\tilde{\nu}_3}<m_{\tilde{\nu}_{1,2}}$, the amplitudes $A^C_{R(1,1)}$ and $A^C_{R(1,2)}$, 
$A^C_{R(2,1)}$ and $A^C_{R(2,2)}$ cancel out. 
As a result, $A_R^C \simeq A^C_{R(1,3)}+A^C_{R(2,3)}$: one obtains a small amplitude 
contributing to the muon decay.

%
%%%%%%%%%%%%%%%%%%%%%%%%%%%%%%%%%%%%%%%%%%%%%%%%%%%%%%

%

\begin{thebibliography}{99}
%%%%%%%%%%%%%%%%%%%%%%%%%%%%%%%%%%%%%%%%%%%%%%%%%%%%%%
\small
%
\bibitem{Canepa:2019hph} A.~Canepa,
  %``Searches for Supersymmetry at the Large Hadron Collider,''
  Rev.\ Phys.\  {\bf 4} (2019) 100033.
  doi:10.1016/j.revip.2019.100033
%
\bibitem{witten}  E.~Witten, %``Dynamical Breaking of Supersymmetry,'' \
  Nucl.\ Phys.\ B {\bf 188}, 513 (1981); R.~K.~Kaul, %``Gauge Hierarchy
 % in a Supersymmetric Model,'' \ 
Phys.\ Lett.\ B {\bf 109}, 19 (1982).
%
\bibitem{goldberg} H.~Goldberg,
  %``Constraint on the Photino Mass from Cosmology,''
  Phys.\ Rev.\ Lett.\  {\bf 50} (1983) 1419;
J.~R.~Ellis, J.~S.~Hagelin, D.~V.~Nanopoulos, K.~A.~Olive and M.~Srednicki,
  %``Supersymmetric Relics from the Big Bang,''
  Nucl.\ Phys.\ B {\bf 238} (1984) 453.
%
\bibitem{SUSYnus} J.~A.~Casas and A.~Ibarra,
  %``Oscillating neutrinos and muon ---> e, gamma,''
  Nucl.\ Phys.\ B {\bf 618} (2001) 171; 
F.~Deppisch, H.~Pas, A.~Redelbach, R.~Ruckl and Y.~Shimizu,
  %``Neutrino oscillations, SUSY seesaw mechanism and charged lepton flavor violation,''
  Nucl.\ Phys.\ Proc.\ Suppl.\  {\bf 116} (2003) 316.
%
\bibitem{wss} H.~Baer and X.~Tata,
  %``Weak scale supersymmetry: From superfields to scattering events,''
  Cambridge, UK: Univ. Pr. (2006) 537 p.;
M.~Drees, R.~Godbole and P.~Roy,
  %``Theory and phenomenology of sparticles: An account of four-dimensional N=1 supersymmetry in high energy physics,''
  Hackensack, USA: World Scientific (2004) 555 p;
S.~P.~Martin,
  %``A Supersymmetry primer,''
  Adv.\ Ser.\ Direct.\ High Energy Phys.\  {\bf 21} (2010) 1,
  [hep-ph/9709356];
D.~J.~H.~Chung, L.~L.~Everett, G.~L.~Kane, S.~F.~King, J.~D.~Lykken and L.~T.~Wang,
  %``The Soft supersymmetry breaking Lagrangian: Theory and applications,''
  Phys.\ Rept.\  {\bf 407} (2005) 1.
%
\bibitem{gauge} S.~Dimopoulos, S.~Raby and F.~Wilczek,
  %``Supersymmetry and the Scale of Unification,''
  Phys.\ Rev.\ D {\bf 24} (1981) 1681; 
U.~Amaldi, W.~de Boer and H.~Furstenau,
  %``Comparison of grand unified theories with electroweak and strong coupling constants measured at LEP,''
  Phys.\ Lett.\ B {\bf 260}, 447 (1991);
J.~R.~Ellis, S.~Kelley and D.~V.~Nanopoulos,
  %``Probing the desert using gauge coupling unification,''
  Phys.\ Lett.\ B {\bf 260} (1991) 131;
P.~Langacker and M.~x.~Luo,
  %``Implications of precision electroweak experiments for $M_t$, $\rho_{0}$, $\sin^2\theta_W$ and grand unification,''
  Phys.\ Rev.\ D {\bf 44} (1991) 817.
%
\bibitem{rewsb} L.~E.~Ibanez and G.~G.~Ross,
  %``SU(2)-L x U(1) Symmetry Breaking as a Radiative Effect of Supersymmetry Breaking in Guts,''
  Phys.\ Lett.\  {\bf 110B} (1982) 215; 
K. Inoue {\it et al.} Prog. Theor. Phys. {\bf 68}, 927 (1982)
and {\bf 71}, 413 (1984);
L.~Iba\~nez, Phys. Lett. {\bf B118}, 73 (1982);
 H.~P.~Nilles, M.~Srednicki and D.~Wyler,
  %``Weak Interaction Breakdown Induced by Supergravity,''                      
  Phys.\ Lett.\ B {\bf 120} (1983) 346;
J.~Ellis, J.~Hagelin, D.~Nanopoulos and M.~Tamvakis,
Phys. Lett. {\bf B125}, 275 (1983);
L.~Alvarez-Gaum\'e. J.~Polchinski and M.~Wise,
Nucl. Phys. {\bf B221}, 495 (1983);
B.~A.~Ovrut and S.~Raby,
  %``The Locally Supersymmetric Geometrical Hierarchy Model,''
  Phys.\ Lett.\ B {\bf 130} (1983) 277;
for a review, see
L.~E.~Ibanez and G.~G.~Ross,
  %``Supersymmetric Higgs and radiative electroweak breaking,''                 
  Comptes Rendus Physique {\bf 8} (2007) 1013.
%
\bibitem{mhiggs} H.~E.~Haber and R.~Hempfling,
  %``Can the mass of the lightest Higgs boson of the minimal supersymmetric model be larger than m(Z)?,''
  Phys.\ Rev.\ Lett.\  {\bf 66} (1991) 1815;
J.~R.~Ellis, G.~Ridolfi and F.~Zwirner,
  %``Radiative corrections to the masses of supersymmetric Higgs bosons,''
  Phys.\ Lett.\ B {\bf 257} (1991) 83;
Y.~Okada, M.~Yamaguchi and T.~Yanagida,
  %``Upper bound of the lightest Higgs boson mass in the minimal supersymmetric standard model,''
  Prog.\ Theor.\ Phys.\  {\bf 85} (1991) 1;
M.~Carena, M.~Quiros and C.~E.~M.~Wagner,
  %``Effective potential methods and the Higgs mass spectrum in the MSSM,''
  Nucl.\ Phys.\ B {\bf 461} (1996) 407;
V.~Barger, P.~Huang, M.~Ishida and W.~Y.~Keung,
  %``Flavor-tuned 125 GeV supersymmetric Higgs boson at the LHC: Test of minimal and natural supersymmetric models,''
  Phys.\ Rev.\ D {\bf 87} (2013) no.1,  015003;
For a review, see {\it e.g.}  M.~Carena and H.~E.~Haber,
  %``Higgs boson theory and phenomenology,''
  Prog.\ Part.\ Nucl.\ Phys.\  {\bf 50} (2003) 63.
%
\bibitem{sven} S.~Heinemeyer, W.~Hollik, D.~Stockinger, A.~M.~Weber and G.~Weiglein,
  %``Precise prediction for M(W) in the MSSM,''
  JHEP {\bf 0608} (2006) 052.
%
\bibitem{nureview} A.~Baha Balantekin and B.~Kayser,
  %``On the Properties of Neutrinos,''
  Ann.\ Rev.\ Nucl.\ Part.\ Sci.\  {\bf 68} (2018) 313.
%
\bibitem{Capozzi:2017ipn} 
  F.~Capozzi, E.~Di Valentino, E.~Lisi, A.~Marrone, A.~Melchiorri and A.~Palazzo,
  %``Global constraints on absolute neutrino masses and their ordering,''
  Phys.\ Rev.\ D {\bf 95}, no. 9, 096014 (2017)
  doi:10.1103/PhysRevD.95.096014
  [arXiv:1703.04471 [hep-ph]];
I.~Esteban, M.~C.~Gonzalez-Garcia, A.~Hernandez-Cabezudo, M.~Maltoni and T.~Schwetz,
  %``Global analysis of three-flavour neutrino oscillations: synergies and tensions in the determination of $\theta_23, \delta_CP$, and the mass ordering,''
  JHEP {\bf 1901} (2019) 106.
%
\bibitem{Tanabashi:2018oca} 
  M.~Tanabashi {\it et al.} [Particle Data Group],
  %``Review of Particle Physics,''
  Phys.\ Rev.\ D {\bf 98}, no. 3, 030001 (2018).
  doi:10.1103/PhysRevD.98.030001.
%  
\bibitem{Aartsen:2019eht} 
  M.~G.~Aartsen {\it et al.},
  %``Probing the Neutrino Mass Ordering with Atmospheric Neutrinos from Three Years of IceCube DeepCore Data,''
  arXiv:1902.07771 [hep-ex].
%  
\bibitem{NOvA:2018gge} 
  M.~A.~Acero {\it et al.} [NOvA Collaboration],
  %``New constraints on oscillation parameters from $\nu_e$ appearance and $\nu_\mu$ disappearance in the NOvA experiment,''
  Phys.\ Rev.\ D {\bf 98}, 032012 (2018)
  doi:10.1103/PhysRevD.98.032012
  [arXiv:1806.00096 [hep-ex]].
  %
\bibitem{Deppisch:2018flu} 
  T.~Deppisch, S.~Schacht and M.~Spinrath,
  %``Confronting SUSY SO(10) with updated Lattice and Neutrino Data,''
  JHEP {\bf 1901}, 005 (2019)
  doi:10.1007/JHEP01(2019)005
  [arXiv:1811.02895 [hep-ph]].
  %
\bibitem{cmssm} For a review, see {\it e.g.}
R.~Arnowitt and P.~Nath,
  %``Developments in Supergravity Unified Models,''                                                   
  In *Kane, G.L. (ed.): Perspectives on supersymmetry II* 222-243
  [arXiv:0912.2273 [hep-ph]]  and references therein;
V.~D.~Barger, M.~S.~Berger and P.~Ohmann,
  %``Supersymmetric grand unified theories: Two loop evolution of gauge and Yukawa couplings,''
  Phys.\ Rev.\ D {\bf 47} (1993) 1093 and 
  Phys.\ Rev.\ D {\bf 49} (1994) 4908;
G.~L.~Kane, C.~F.~Kolda, L.~Roszkowski and J.~D.~Wells,
  %``Study of constrained minimal supersymmetry,''
Phys.\ Rev.\ D {\bf 49} (1994) 6173.
%
\bibitem{TheMEG:2016wtm} 
  A.~M.~Baldini {\it et al.} [MEG Collaboration],
  %``Search for the lepton flavour violating decay $\mu ^+ \rightarrow \mathrm {e}^+ \gamma $ with the full dataset of the MEG experiment,''
  Eur.\ Phys.\ J.\ C {\bf 76}, no. 8, 434 (2016)
  doi:10.1140/epjc/s10052-016-4271-x
  [arXiv:1605.05081 [hep-ex]].
  %
\bibitem{Renga:2018fpd} 
  F.~Renga [MEG Collaboration],
  %``The quest for $\mu \to e \gamma$: present and future,''
  Hyperfine Interact.\  {\bf 239}, no. 1, 58 (2018)
  doi:10.1007/s10751-018-1534-y
  [arXiv:1811.05921 [hep-ex]].
  %
  \bibitem{Calibbi:2012gr} 
  L.~Calibbi, D.~Chowdhury, A.~Masiero, K.~M.~Patel and S.~K.~Vempati,
  %``Status of supersymmetric type-I seesaw in SO(10) inspired models,''
  JHEP {\bf 1211}, 040 (2012)
  doi:10.1007/JHEP11(2012)040
  [arXiv:1207.7227 [hep-ph]].
%
  \bibitem{Bora:2014mna} 
  K.~Bora and G.~Ghosh,
  %``Charged lepton flavor violation $\mu \rightarrow e \gamma $ in $\mu $ – $ \tau $ symmetric SUSY SO(10) mSUGRA, NUHM, NUGM, and NUSM theories and LHC,''
  Eur.\ Phys.\ J.\ C {\bf 75}, no. 9, 428 (2015)
  doi:10.1140/epjc/s10052-015-3617-0
  [arXiv:1410.1265 [hep-ph]].
%
\bibitem{nuhm2}  D.~Matalliotakis and H.~P.~Nilles,
  %``Implications of nonuniversality of soft terms in supersymmetric grand unified theories,''
  Nucl.\ Phys.\ B {\bf 435} (1995) 115;
M.~Olechowski and S.~Pokorski,
  %``Electroweak symmetry breaking with nonuniversal scalar soft terms and large tan beta solutions,''
  Phys.\ Lett.\ B {\bf 344} (1995) 201;
P.~Nath and R.~L.~Arnowitt,
  %``Nonuniversal soft SUSY breaking and dark matter,''
  Phys.\ Rev.\ D {\bf 56} (1997) 2820;
J. Ellis, K. Olive and Y. Santoso, Phys. Lett. {\bf B539} (2002) 107;
J. Ellis, T. Falk, K. Olive and Y. Santoso, 
Nucl. Phys. {\bf B652} (2003) 259;
H.~Baer, A.~Mustafayev, S.~Profumo, A.~Belyaev and X. Tata, 
JHEP{\bf 0507} (2005) 065.
%
\bibitem{Baer:2012up} 
  H.~Baer, V.~Barger, P.~Huang, A.~Mustafayev and X.~Tata,
  %``Radiative natural SUSY with a 125 GeV Higgs boson,''
  Phys.\ Rev.\ Lett.\  {\bf 109}, 161802 (2012)
  doi:10.1103/PhysRevLett.109.161802
  [arXiv:1207.3343 [hep-ph]].
%
\bibitem{Baer:2012cf}
  H.~Baer, V.~Barger, P.~Huang, D.~Mickelson, A.~Mustafayev and X.~Tata,
  %``Radiative natural supersymmetry: Reconciling electroweak fine-tuning and the Higgs boson mass,''
  Phys.\ Rev.\ D {\bf 87} (2013) no.11,  115028
  doi:10.1103/PhysRevD.87.115028
  [arXiv:1212.2655 [hep-ph]].
%
\bibitem{Baer:2018hpb} 
  H.~Baer, V.~Barger, J.~S.~Gainer, D.~Sengupta, H.~Serce and X.~Tata,
  %``LHC luminosity and energy upgrades confront natural supersymmetry models,''
  Phys.\ Rev.\ D {\bf 98}, no. 7, 075010 (2018)
  doi:10.1103/PhysRevD.98.075010
%
\bibitem{nAMSB} H.~Baer, V.~Barger and D.~Sengupta,
  %``Anomaly mediated SUSY breaking model retrofitted for naturalness,''
  Phys.\ Rev.\ D {\bf 98} (2018) no.1,  015039.
%
\bibitem{CidVidal:2018eel} 
  X.~Cid Vidal  {\it et al.},
  %``Beyond the Standard Model Physics at the HL-LHC and HE-LHC,''
  arXiv:1812.07831 [hep-ph].
%  
\bibitem{Baer:2017uvn}
  H.~Baer, V.~Barger, H.~Serce and K.~Sinha,
  %``Higgs and superparticle mass predictions from the landscape,''
  JHEP {\bf 1803} (2018) 002
  doi:10.1007/JHEP03(2018)002
  [arXiv:1712.01399 [hep-ph]].
%
\bibitem{Baer:2016ucr} 
  H.~Baer, V.~Barger and H.~Serce,
  %``SUSY under siege from direct and indirect WIMP detection experiments,''
  Phys.\ Rev.\ D {\bf 94}, no. 11, 115019 (2016)
  doi:10.1103/PhysRevD.94.115019
  [arXiv:1609.06735 [hep-ph]].
  %
\bibitem{boltz} 
  K.~J.~Bae, H.~Baer, A.~Lessa and H.~Serce,
  %``Coupled Boltzmann computation of mixed axion neutralino dark matter in the SUSY DFSZ axion model,''
  JCAP {\bf 1410}, no. 10, 082 (2014)
  doi:10.1088/1475-7516/2014/10;
    K.~J.~Bae, H.~Baer, A.~Lessa and H.~Serce,
  %``Mixed axion-wino dark matter,''
  Front.\ in Phys.\  {\bf 3}, 49 (2015)
  doi:10.3389/fphy.2015.00049
  [arXiv:1502.07198 [hep-ph]];
   H.~Baer, A.~Lessa, S.~Rajagopalan and W.~Sreethawong,
  %``Mixed axion/neutralino cold dark matter in supersymmetric models,''
  JCAP {\bf 1106}, 031 (2011)
  doi:10.1088/1475-7516/2011/06/031
  [arXiv:1103.5413 [hep-ph]].
%  
  \bibitem{Serce:2017vtk} 
  H.~Serce,
  %``Implications of Mixed Axion-Neutralino Dark Matter,''
  AIP Conf.\ Proc.\  {\bf 1900}, no. 1, 040007 (2017)
  doi:10.1063/1.5010125
  [arXiv:1703.10151 [hep-ph]];
  K.~J.~Bae, H.~Baer and H.~Serce,
  %``Prospects for axion detection in natural SUSY with mixed axion-higgsino dark matter: back to invisible?,''
  JCAP {\bf 1706}, no. 06, 024 (2017)
  doi:10.1088/1475-7516/2017/06/024
  [arXiv:1705.01134 [hep-ph]].
  %
\bibitem{seesaw} H.~Fritzsch and P.~Minkowski,
  %``Vector-Like Weak Currents, Massive Neutrinos, and Neutrino Beam Oscillations,''
  Phys.\ Lett.\  {\bf 62B} (1976) 72.
  doi:10.1016/0370-2693(76)90051-4;
P.~Minkowski,
  %``$\mu \to e\gamma$ at a Rate of One Out of $10^{9}$ Muon Decays?,''
  Phys.\ Lett.\  {\bf 67B} (1977) 421.
  doi:10.1016/0370-2693(77)90435-X;
M.~Gell-Mann, P.~Ramond and R.~Slansky,
  %``Complex Spinors and Unified Theories,''
  Conf.\ Proc.\ C {\bf 790927} (1979) 315
  [arXiv:1306.4669 [hep-th]];
T.~Yanagida,
  %``Horizontal gauge symmetry and masses of neutrinos,''
  Conf.\ Proc.\ C {\bf 7902131} (1979) 95;
R.~N.~Mohapatra and G.~Senjanovic,
  %``Neutrino Mass and Spontaneous Parity Nonconservation,''
  Phys.\ Rev.\ Lett.\  {\bf 44} (1980) 912, 
  doi:10.1103/PhysRevLett.44.912 .
%
\bibitem{Hisano} 
  J.~Hisano, T.~Moroi, K.~Tobe and M.~Yamaguchi,
  %``Lepton flavor violation via right-handed neutrino Yukawa couplings in supersymmetric standard model,''
  Phys.\ Rev.\ D {\bf 53}, 2442 (1996)
  doi:10.1103/PhysRevD.53.2442
  [hep-ph/9510309];
    J.~Hisano and D.~Nomura,
  %``Solar and atmospheric neutrino oscillations and lepton flavor violation in supersymmetric models with the right-handed neutrinos,''
  Phys.\ Rev.\ D {\bf 59}, 116005 (1999)
  doi:10.1103/PhysRevD.59.116005
  [hep-ph/9810479].
%
\bibitem{Masiero} 
  A.~Masiero, S.~K.~Vempati and O.~Vives,
  %``Seesaw and lepton flavor violation in SUSY SO(10),''
  Nucl.\ Phys.\ B {\bf 649}, 189 (2003)
  doi:10.1016/S0550-3213(02)01031-3
  [hep-ph/0209303];
    L.~Calibbi, A.~Faccia, A.~Masiero and S.~K.~Vempati,
  %``Lepton flavour violation from SUSY-GUTs: Where do we stand for MEG, PRISM/PRIME and a super flavour factory,''
  Phys.\ Rev.\ D {\bf 74}, 116002 (2006)
  doi:10.1103/PhysRevD.74.116002
  [hep-ph/0605139];
    L.~Calibbi, D.~Chowdhury, A.~Masiero, K.~M.~Patel and S.~K.~Vempati,
  %``Status of supersymmetric type-I seesaw in SO(10) inspired models,''
  JHEP {\bf 1211}, 040 (2012)
  doi:10.1007/JHEP11(2012)040
  [arXiv:1207.7227 [hep-ph]].
%
\bibitem{Moroi:2013vya} 
  T.~Moroi, M.~Nagai and T.~T.~Yanagida,
  %``Lepton Flavor Violations in High-Scale SUSY with Right-Handed Neutrinos,''
  Phys.\ Lett.\ B {\bf 728}, 342 (2014)
  doi:10.1016/j.physletb.2013.11.058
  [arXiv:1305.7357 [hep-ph]].
%
\bibitem{Hirsch:2012ti} 
  M.~Hirsch, F.~R.~Joaquim and A.~Vicente,
  %``Constrained SUSY seesaws with a 125 GeV Higgs,''
  JHEP {\bf 1211}, 105 (2012)
  doi:10.1007/JHEP11(2012)105
  [arXiv:1207.6635 [hep-ph]].
  %
\bibitem{Barger:2009gc}
  V.~Barger, D.~Marfatia, A.~Mustafayev and A.~Soleimani,
  %``SUSY dark matter and lepton flavor violation,''
  Phys.\ Rev.\ D {\bf 80} (2009) 076004
  doi:10.1103/PhysRevD.80.076004
  [arXiv:0908.0941 [hep-ph]].
%
\bibitem{Borzumati:1986qx}
  F.~Borzumati and A.~Masiero,
  %``Large Muon and electron Number Violations in Supergravity Theories,''
  Phys.\ Rev.\ Lett.\  {\bf 57} (1986) 961.
%
\bibitem{Vafa:2005ui}
  C.~Vafa,
  %``The String landscape and the swampland,''
  hep-th/0509212.
%
\bibitem{Halverson:2018xge}
  J.~Halverson and P.~Langacker,
  %``TASI Lectures on Remnants from the String Landscape,''
  PoS TASI {\bf 2017} (2018) 019
  doi:10.22323/1.305.0019
  [arXiv:1801.03503 [hep-th]].
%
\bibitem{Buchmuller:2007zd}
  W.~Buchmuller, K.~Hamaguchi, O.~Lebedev, S.~Ramos-Sanchez and M.~Ratz,
  %``Seesaw neutrinos from the heterotic string,''
  Phys.\ Rev.\ Lett.\  {\bf 99} (2007) 021601
  doi:10.1103/PhysRevLett.99.021601
  [hep-ph/0703078 [HEP-PH]].
%
\bibitem{Buchmuller:2005sh}
  W.~Buchmuller, K.~Hamaguchi, O.~Lebedev and M.~Ratz,
  %``Local grand unification,''
  hep-ph/0512326.
%
\bibitem{Antusch:2005gp} 
S.~Antusch, J.~Kersten, M.~Lindner, M.~Ratz and M.~A.~Schmidt,
  %``Running neutrino mass parameters in see-saw scenarios,''
  JHEP {\bf 0503}, 024 (2005)
  doi:10.1088/1126-6708/2005/03/024
  [hep-ph/0501272].
%  
  \bibitem{Petcov:2003zb} 
  S.~T.~Petcov, S.~Profumo, Y.~Takanishi and C.~E.~Yaguna,
  %``Charged lepton flavor violating decays: Leading logarithmic approximation versus full RG results,''
  Nucl.\ Phys.\ B {\bf 676}, 453 (2004)
  doi:10.1016/j.nuclphysb.2003.10.020
  [hep-ph/0306195].
%
\bibitem{Baldini:2018nnn} 
  A.~M.~Baldini {\it et al.} [MEG II Collaboration],
  %``The design of the MEG II experiment,''
  Eur.\ Phys.\ J.\ C {\bf 78}, no. 5, 380 (2018)
  doi:10.1140/epjc/s10052-018-5845-6
  [arXiv:1801.04688 [physics.ins-det]].
  %
\bibitem{Aubert:2009ag} 
  B.~Aubert {\it et al.} [BaBar Collaboration],
  %``Searches for Lepton Flavor Violation in the Decays tau+- ---> e+- gamma and tau+- ---> mu+- gamma,''
  Phys.\ Rev.\ Lett.\  {\bf 104}, 021802 (2010)
  doi:10.1103/PhysRevLett.104.021802
  [arXiv:0908.2381 [hep-ex]].
  %
  \bibitem{BelleII} 
  K.~Hayasaka [Belle Collaboration],
  %``Recent LFV results on tau lepton from Belle,''
  Nucl.\ Phys.\ Proc.\ Suppl.\  {\bf 225-227}, 169 (2012).
  doi:10.1016/j.nuclphysbps.2012.02.036;
  P.~Branchini [Belle-II Collaboration],
  %``The Belle II Experiment: Status and Prospects,''
  Universe {\bf 4}, no. 10, 101 (2018).
  %
  \bibitem{Arganda:2005ji} 
  E.~Arganda and M.~J.~Herrero,
  %``Testing supersymmetry with lepton flavor violating tau and mu decays,''
  Phys.\ Rev.\ D {\bf 73}, 055003 (2006)
  doi:10.1103/PhysRevD.73.055003
  [hep-ph/0510405].
  %
  \bibitem{Bellgardt:1987du} 
  U.~Bellgardt {\it et al.} [SINDRUM Collaboration],
  %``Search for the Decay mu+ ---> e+ e+ e-,''
  Nucl.\ Phys.\ B {\bf 299}, 1 (1988).
  %
  \bibitem{Perrevoort:2018ttp} 
  A.~K.~Perrevoort [Mu3e Collaboration],
  %``The Rare and Forbidden: Testing Physics Beyond the Standard Model with Mu3e,''
  SciPost Phys.\ Proc.\  {\bf 1}, 052 (2019)
  doi:10.21468/SciPostPhysProc.1.052
  [arXiv:1812.00741 [hep-ex]].
  %
\bibitem{Kitano:2002mt} 
  R.~Kitano, M.~Koike and Y.~Okada,
  %``Detailed calculation of lepton flavor violating muon electron conversion rate for various nuclei,''
  Phys.\ Rev.\ D {\bf 66}, 096002 (2002)
  Erratum: [Phys.\ Rev.\ D {\bf 76}, 059902 (2007)]
  doi:10.1103/PhysRevD.76.059902, 10.1103/PhysRevD.66.096002
  [hep-ph/0203110].
  %
  \bibitem{Bartoszek:2014mya} 
  L.~Bartoszek {\it et al.} [Mu2e Collaboration],
  %``Mu2e Technical Design Report,''
  arXiv:1501.05241 [physics.ins-det].
  %
\bibitem{Nilles:2014owa}
  H.~P.~Nilles and P.~K.~S.~Vaudrevange,
  %``Geography of Fields in Extra Dimensions: String Theory Lessons for Particle Physics,''
  Mod.\ Phys.\ Lett.\ A {\bf 30} (2015) no.10,  1530008
  doi:10.1142/S0217732315300086
  [arXiv:1403.1597 [hep-th]].
%
\bibitem{Paige:2003mg} 
  F.~E.~Paige, S.~D.~Protopopescu, H.~Baer and X.~Tata,
  %``ISAJET 7.69: A Monte Carlo event generator for pp, anti-p p, and e+e- reactions,''
  hep-ph/0312045.
%
\bibitem{Chowdhury:2011zr} 
  D.~Chowdhury, R.~Garani and S.~K.~Vempati,
  %``SUSEFLAV: Program for supersymmetric mass spectra with seesaw mechanism and rare lepton flavor violating decays,''
  Comput.\ Phys.\ Commun.\  {\bf 184}, 899 (2013)
  doi:10.1016/j.cpc.2012.10.031
  [arXiv:1109.3551 [hep-ph]].
%
\bibitem{LEP2} 
  LEPSUSYWG, ALEPH, DELPHI, L3 and OPAL experiments, note LEPSUSYWG/yy-nn
  (http://lepsusy.web.cern.ch/lepsusy/Welcome.html).
  %
\bibitem{Endo:2015oia} 
  M.~Endo, T.~Moroi and M.~M.~Nojiri,
  %``Footprints of Supersymmetry on Higgs Decay,''
  JHEP {\bf 1504}, 176 (2015)
  doi:10.1007/JHEP04(2015)176
  [arXiv:1502.03959 [hep-ph]].
  %
\bibitem{Witten:1996mz} 
  E.~Witten,
  %``Strong coupling expansion of Calabi-Yau compactification,''
  Nucl.\ Phys.\ B {\bf 471}, 135 (1996)
  doi:10.1016/0550-3213(96)00190-3
  [hep-th/9602070].
%    
\bibitem{Arana-Catania:2013xma} 
  M.~Arana-Catania, E.~Arganda and M.~J.~Herrero,
  %``Non-decoupling SUSY in LFV Higgs decays: a window to new physics at the LHC,''
  JHEP {\bf 1309}, 160 (2013)
  Erratum: [JHEP {\bf 1510}, 192 (2015)]
  doi:10.1007/JHEP10(2015)192, 10.1007/JHEP09(2013)160
  [arXiv:1304.3371 [hep-ph]].
%  
  \bibitem{Bernstein:2019fyh} 
  R.~H.~Bernstein [Mu2e Collaboration],
  %``The Mu2e Experiment,''
  Front.\ in Phys.\  {\bf 7}, 1 (2019)
  doi:10.3389/fphy.2019.00001
  [arXiv:1901.11099 [physics.ins-det]].
  %
  \bibitem{Cirigliano:2009bz} 
  V.~Cirigliano, R.~Kitano, Y.~Okada and P.~Tuzon,
  %``On the model discriminating power of mu ---> e conversion in nuclei,''
  Phys.\ Rev.\ D {\bf 80}, 013002 (2009)
  doi:10.1103/PhysRevD.80.013002
  [arXiv:0904.0957 [hep-ph]].
  %
\bibitem{Abusalma:2018xem} 
  F.~Abusalma {\it et al.} [Mu2e Collaboration],
  %``Expression of Interest for Evolution of the Mu2e Experiment,''
  arXiv:1802.02599 [physics.ins-det].
  %
    \bibitem{Binosi:2008ig} 
  D.~Binosi, J.~Collins, C.~Kaufhold and L.~Theussl,
  %``JaxoDraw: A Graphical user interface for drawing Feynman diagrams. Version 2.0 release notes,''
  Comput.\ Phys.\ Commun.\  {\bf 180}, 1709 (2009)
  doi:10.1016/j.cpc.2009.02.020
  [arXiv:0811.4113 [hep-ph]].
  %
\end{thebibliography}
\end{document}